\documentclass[11pt]{article}
\usepackage{xcolor}
\usepackage{ulem}
\usepackage{dcolumn}
\usepackage{bm}
\usepackage{amsfonts}
\usepackage{amsbsy}
\usepackage{graphicx}
\usepackage{psfrag}
\usepackage{verbatim}
\usepackage{color}
\usepackage{amsmath}
\usepackage{epsfig,bm,epsf,float}
\usepackage{mathrsfs}
\usepackage{latexsym}
\usepackage{mathrsfs}
\usepackage{soul}
\usepackage{cancel}
\usepackage{newtxtext}
\usepackage{newtxmath}
\usepackage{natbib}
\usepackage{hyperref}
\usepackage{authblk}
\usepackage{blindtext}
\usepackage[a4paper, total={6.3in, 10in}]{geometry}
\usepackage[stdsubgroups]{nomencl} % Nomenclature
\makenomenclature

\usepackage{ifthen}
\renewcommand{\nomgroup}[1]{%
\ifthenelse{\equal{#1}{G}}{\item[\textbf{Greek Symbols}]}{%
\ifthenelse{\equal{#1}{S}}{\item[\textbf{Subscripts}]}{}}}

\graphicspath{{./fig/}}
\usepackage{caption}
\usepackage{subcaption}
\newcommand{\Reyn}{\text{\textit{Re}}}
\newcommand{\K}{\mathcal{K}}
\newcommand{\ie}{i.e.\ }
\newcommand{\vect}[1]{\boldsymbol{#1}}

\newcommand{\ue}{\mathrm{e}}
\newcommand{\ui}{\mathrm{i}}

\newcommand{\ud}{\mathrm{d}}

\newcommand{\ms}{\kern.10em\relax}
\newcommand{\pfi}[2]{\ensuremath{{\partial #1}/{\partial #2}}}
\newcommand{\D}{\mathcal{D}}
\newcommand{\DD}{\mathcal{D}^2}
\newcommand{\Pils}{{\Pi_\ell^\ast}}
\newcommand{\omegas}{{\omega^\ast}}
\newcommand{\Us}{{U^\ast}}

\newcommand{\Hs}{{H^\ast}}
\newcommand{\Ks}{{\K^\ast}}
\newcommand{\ts}{{t^\ast}}
\newcommand{\xs}{{x^\ast}}
\newcommand{\ys}{{y^\ast}}
\newcommand{\ps}{{p^\ast}}

\newcommand{\hs}{{h^\ast}}
\newcommand{\ub}{\bar{u}}
\newcommand{\vb}{\bar{v}}
\newcommand{\pb}{\bar{p}}
\newcommand{\hb}{\bar{h}}
\newcommand{\uh}{\skew3\hat{u}}
\newcommand{\vh}{\skew3\hat{v}}
\newcommand{\ph}{\skew3\hat{p}}
\newcommand{\hh}{\skew3\hat{h}}

\newcommand{\Real}{\mathfrak{Re}}
\author[1,2]{Antonio J. Bárcenas-Luque\thanks{ajbarcenas@ugr.es}}
\author[3]{W. Coenen}
\author[4,5]{C. Gutiérrez-Montes}
\author[1,2]{C. Martínez-Bazán}
\affil[1]{Departamento de Mecánica de Estructuras e Ingeniería Hidráulica, Universidad de Granada, Campus Fuentenueva s/n, 18071 Granada, Spain.}
\affil[2]{Andalusian Institute for Earth System Research, University of Granada, Avda. del Mediterráneo s/n, 18006 Granada, Spain.}
\affil[3]{Grupo de Mecánica de Fluidos, Departamento de Ingeniería Térmica y de Fluidos, Universidad Carlos III de Madrid, Spain.}
\affil[4]{\'Area de Mecánica de Fluidos, Departamento de Ingeniería Mecánica y Minera. Universidad de Jaén. Campus de las Lagunillas, 23071 Jaén, Spain.}
\affil[5]{Andalusian Institute for Earth System Research, Universidad de Jaén, Campus de las Lagunillas, Jaén, 23071, Spain.}
\date{}

\begin{document}
\normalem
\title{Floquet stability analysis of a two-layer oscillatory flow near a flexible wall}

\maketitle

%%%%%%%%%%%%%%%%%%%%%%%%%%%%%%%%%%%%%%%%%%%%%%%%%%%%%%%%%%%%%%%%%%%%%%%%%%%%%%%%%%%%%%%%%%%%%%%%%%%%%%%%%%%%%%%%%%%%%%%%
%%%%%%%%%%%%%%%%%%%%%%%%%%%%%%%%%%%%%%%%%%%%%%%%%%%%%%%%%%%%%%%%%%%%%%%%%%%%%%%%%%%%%%%%%%%%%%%%%%%%%%%%%%%%%%%%%%%%%%%%
%%%%%%%%%%%%%%%%%%%%%%%%%%%%%%%%%%%%%%%%%%%%%%%%%%%%%%%%%%%%%%%%%%%%%%%%%%%%%%%%%%%%%%%%%%%%%%%%%%%%%%%%%%%%%%%%%%%%%%%%

\begin{abstract}
We investigate the linear Floquet stability of two fluid layers undergoing oscillations in the direction parallel to
the flexible wall that separates them. This canonical configuration is inspired by the cerebrospinal fluid flow in the
spinal canal of subjects with hydro-/syringomyelia.
The analysis focuses on the marginal conditions for the onset of instability, and how these depend on the
spatial wavelength of the perturbation, and on the values of the control parameters, which are the two channel
widths, the Reynolds number, and the wall stiffness. Unstable perturbations are found to oscillate synchronous with the
base flow. The wavelength of the most unstable perturbation,
of the order of the stroke length of the basic oscillatory motion, depends strongly on the wall stiffness,
but is only weakly influenced by the channel widths and the Reynolds number. In general, around criticality, it was
found that increasing the Reynolds number has a destabilizing effect, and that decreasing the canal widths stabilizes
the instability. The wall stiffness on the other hand has a non-monotonic effect, exhibiting an intermediate value for
which the instability is maximally amplified. The present analysis is a first step towards a better understanding of the
physical mechanisms that govern many (bio)fluid mechanical problems that involve oscillatory flows near compliant walls.
\end{abstract}

%%%%%%%%%%%%%%%%%%%%%%%%%%%%%%%%%%%%%%%%%%%%%%%%%%%%%%%%%%%%%%%%%%%%%%%%%%%%%%%%%%%%%%%%%%%%%%%%%%%%%%%%%%%%%%%%%%%%%%%%
%%%%%%%%%%%%%%%%%%%%%%%%%%%%%%%%%%%%%%%%%%%%%%%%%%%%%%%%%%%%%%%%%%%%%%%%%%%%%%%%%%%%%%%%%%%%%%%%%%%%%%%%%%%%%%%%%%%%%%%%
%%%%%%%%%%%%%%%%%%%%%%%%%%%%%%%%%%%%%%%%%%%%%%%%%%%%%%%%%%%%%%%%%%%%%%%%%%%%%%%%%%%%%%%%%%%%%%%%%%%%%%%%%%%%%%%%%%%%%%%%

\section{Introduction}
\label{sec:intro}

\begin{figure}
    \centering
    \includegraphics[width=1.0\textwidth]{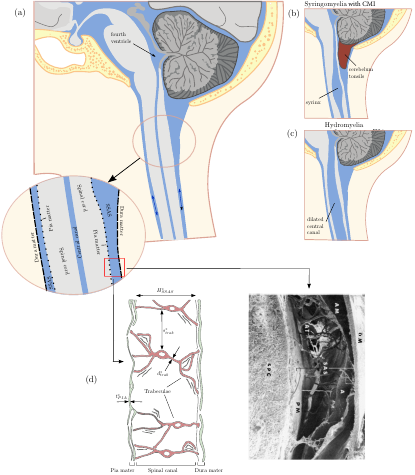}
    \caption{(a) Schematic overview of the main features of the anatomy of the cervical spine. b) Schematic representation of syringomyelia associated with Type I Chiari Malformation. c) Schematic representation of hydromyelia. d) Diagram and photograph \cite{cloyd1974scanning} of the micro-anatomy inside the spinal subarachnoid space; here AM: arachnoid mater, AT: arachnoid trabeculae, A: artery, SAS: subarachnoid space, DM: dura mater, PM: pia Mater, SPC: spinal cord.}
    \label{fig:intro}
\end{figure}

Understanding the role played by pulsatile or oscillatory flows in the vicinity of deformable walls is crucial in many engineering \cite{Jaffrin1971} and biomedical applications \cite{Grotberg2004}. In fact, many of the physiological systems of the human body present this kind of motion. The flow of air in the respiratory system, that of blood in the circulatory system, or the flow of cerebrospinal fluid (CSF) in the central nervous system (CNS), constitute relevant examples where the oscillating driving mechanisms, together with the interaction of the fluid with the compliant walls, determine the nature of the fluid motion \cite{Linninger.etal.2016, Heil2011}. Nevertheless, despite their relevance, many aspects of oscillatory flow in the presence of compliant walls are still poorly understood. In the present paper we focus on the flow of CSF in the cervical spinal subarachnoid space and its interaction with the fluid in the central canal of the spinal cord through the deformation of the spinal cord tissue that separates both fluid layers (Figure~\ref{fig:intro}a). We focus on this configuration in the context of two pathologies, named \emph{syringomyelia} and \emph{hydromyelia}. Both conditions are similar and involve the accumulation of fluid in the spinal cord. In syringomyelia it occurs in the form of a slender fluid-filled cavity, called syrinx, approximately 2-6 cm long and 3-6 mm wide (Figure~\ref{fig:intro}b)\cite{Elliott.etal.2013}, whereas in hydromyelia there is an abnormal dilation of the central canal that typically spans a longer section of the spine, about 3-4 vertebrae (Figure~\ref{fig:intro}c)\cite{roser2010defining}. In hydromyelia, the dilated central canal is considered to be connected to the fourth ventricle (which is the case in newborns and young children), whereas in syringomyelia, the cavity may be isolated as a result of age-related obstruction of the central canal. Nevertheless, the distinction between both diseases is often hard to make, and some physicians might even use both terms interchangeably. The pathogenesis of these diseases is still unknown, but there is a consensus that CSF hydrodynamics may play an important role. In fact, many cases of hydro-/syringomyelia are associated with a malformation of the hindbrain named `Type I Chiari Malformation' (CMI), in which the cerebellar tonsils descend into the entrance of the spinal subarachnoid space and partially block the natural oscillatory motion of CSF (Figure~\ref{fig:intro}b). This back-and-forth motion of CSF in the spinal subarachnoid space surrounding the spinal cord is in principle coupled to the motion of fluid in the central canal, with associated periodic deformations of the compliant spinal cord tissue. Whereas in healthy subjects these deformations may remain small, in subjects with Chiari Malformation the changed hydrodynamics in the subarachnoid space may induce larger self-sustained deformations of the spinal cord tissue separating the two fluid layers. Prolonged exposure to such deformations may then in turn be associated with additional oedema or fluid transport into the fluid cavity, leading to the growth of cavities, whose size may be marked by the wavelength of the preferential deformations. As a first step to explore if this proposed mechanism is viable, in the present paper we study the underlying fluid-structure interaction problem using a simplified model description.

% Esto decía mas om menos lo mismo:
%The main question that comes to mind is to know if there are self-sustained deformations of the wall that separates both channels due to the movement of the CSF, as well as to know if said deformations together with the wavelengths associated with the most unstable mode are somehow linked to the development of pathologies of the central nervous system related to the central canal, such as syringomyelia or hydromyelia.
%A possible explanation for the development of a syrinx could be based on the fact that starting from a configuration without said pathology and with disturbances of the flow coming from the blockage produced in the SSAS by Chiari Malformation Type 1 (CMT1) (as seen in figure \ref{fig:intro}), the wavelengths associated with the most unstable mode mark in some way the size of the syrinxes. Prolonged exposure to such preferential deformations could produce permanent deformations in the central canal and the formation of slender cavities of a size comparable to that of the wavelengths associated with the most energetic unstable mode.

The main features of the anatomy of the cervical spine that are relevant to the problem at hand are given in figure~\ref{fig:intro}. The spinal subarachnoid space is a thin annular canal, approximately $\Hs_\text{SSAS} \sim 1-4~\text{mm}$ wide, bounded internally by the pia membrane (thickness $t^\ast_\text{pia} \sim 100 \, \mu\text{m}$) and externally by the dura membrane. At the center of the spinal cord lies the central canal, which in healthy subjects has a cross-sectional diameter of approximately $0.5-2$~mm. In the presence of hydro-/syringomyelia its diameter increases to approximately $3-4$~mm. CSF is a water-like fluid with a density $\rho_\text{CSF} \simeq 1000\,\text{kg}/\text{m}^3$ and a kinematic viscosity $\nu_\text{CSF} \simeq 0.7\times10^{-6}\,\text{m}^2/\text{s}$. Its motion in the subarachnoid space is driven mainly by the intracranial pressure fluctuations that occur with each heartbeat as a result of the cyclic variation of the cerebrovascular blood volume (with frequencies on the order of $\omega^{\ast}_\text{CSF}/(2\pi) \sim 1\,\text{Hz}$), resulting in a volume of CSF of approximately $1~\text{ml}$ that is cyclically pushed into and out of the spinal canal, as measured in multiple studies using magnetic resonance imaging \citep[see, for example][]{coenen2019subject, sincomb2022one}. The back-and-forth motion of CSF has an typical stroke length $l^\ast_\text{CSF} \sim 5-10\,\text{mm}$, and a characteristic velocity $U^\ast_\text{CSF} \sim 1-2 \,\text{cm}/\text{s}$. The corresponding instantaneous Reynolds number is therefore on the order of $U^\ast_\text{CSF} \Hs_\text{SSAS} / \nu_\text{CSF} \sim 50$, and the corresponding Womersley number is on the order of $\Hs_\text{SSAS}\sqrt{\omega^\ast_\text{CSF}/\nu_\text{CSF}} \sim 8$. Note that in the model problem to be presented below we will base the Reynolds number on the stroke length, so that its corresponding order of magnitude is $U^\ast_\text{CSF} l^\ast_\text{CSF} / \nu_\text{CSF} \sim 150$. When the central canal is connected to the fourth ventricle, the fluid inside it is exposed to the same longitudinal oscillatory pressure gradient driven by the intracranial pressure fluctuations. Velocity measurements inside the central canal of healthy subjects are however not available in the literature because the diameter of the canal is too small for current MRI techniques. Nevertheless, there are a few measurements of flow inside large syringomyelia cysts, showing oscillatory motion in sync with that of CSF in the subarachnoid space surrounding the spinal cord \cite[for example][]{honey2017syringomyelia, luzzi2021pulsatile}. Various micro-anatomical features, such as nerve roots, denticulate ligaments, and trabeculae connect the pia membrane around the spinal cord with the outer dura mater. Among these, the trabeculae (Figure~\ref{fig:intro}d) are especially relevant for the present problem, since they constitute a network of pillar-like structures that maintain the spinal cord in its position inside the spinal canal and in this manner give stiffness to the otherwise relatively flexible spinal cord tissue. The diameters of the trabeculae are on the order of $d^\ast_\text{trab} \sim 10-20\,\mu\text{m}$, and they are spaced approximately $s^\ast_\text{trab} \sim 500\,\mu\text{m}$ apart \cite[]{stockman2006, gupta2009, salerno2020}. In the problem formulation (\S\ref{ssec:form-goveqs}), we will show with an order of magnitude estimation of the governing fluid-structure interaction equation that the stiffness provided by the trabeculae is indeed the dominating effect that balances the normal force exerted by the two fluid layers on the separating spinal cord.

Inspired by the complex biofluid mechanical problem described in the previous paragraph, in the present paper we study a simplified, canonical configuration, consisting of two infinitely long layers of fluid that undergo an oscillatory motion parallel to a thin planar flexible wall that separates them. We investigate whether the separating wall will remain undeformed, the two oscillatory Stokes layers that form on both sides remaining unperturbed, or whether the configuration acquires a different oscillatory state in which the wall moves together with the two layers. We aim to answer this question by means of a Floquet linear stability analysis, exploring the influence of the different governing parameters such as the layer widths, the Reynolds number, and the wall stiffness. The study of fluid-structure interactions in simplified two-layer oscillating flow configurations such as the one presented here may prove helpful to elucidate some of the mechanisms that intervene in the complex, real-life flow described before.

%\begin{figure}
   % \centering
    %\includegraphics[angle=270,width=0.80\textwidth]{fig/syrinx-crop.pdf}
   % \caption{Syringomyelia associated with Chiari malformation modelling.}
   % \label{fig:syrinx}
%\end{figure}

The literature on the stability of time-periodic flows in the presence of compliant walls is limited.
On the contrary, its rigid counterpart has been studied extensively \cite{Davis.1976}.
The stability of the planar Stokes layer near a rigid wall was investigated by \cite{Hall.1978}
and \cite{Blennerhassett.etal.2002}. Time-periodic channel and pipe flows with rigid walls have been
studied by \cite{Yang.etal.1977}, \cite{vonKerczek.1982}, \cite{Thomas.etal.2011}, \cite{Pier.etal.2017},
and \cite{Pier.etal.2021}, using linear Floquet analyses, nonmodal stability analyses, as well fully
nonlinear direct numerical simulations. Nevertheless, the absence of flexible walls limits the implications
of their findings for the problem under consideration here. The hydrodynamic stability of flows near flexible walls has been mainly focused on steady configurations. Canonical examples include the Couette flow past a flexible membrane \cite{Kumaran.Srivatsan.1998, Thaokar.Kumaran.2002, Kumaran.2021}, and the Poiseuille flow between compliant walls \cite{carpenter_garrad_1985, Davies.Carpenter.1997, Lebbal.etal.2022}. The coupling between the wall displacement and fluid velocity provides additional destabilizing mechanisms that enrich the stability characteristics drastically with respect to the rigid analogues of these flows. To the best of our knowledge, there are only a few studies in the literature on the stability of flows that contain both time-periodicity and flexible walls, among which the most relevant for the problem at hand work are the
stability analyses of plane pulsatile channel flow between compliant walls by \cite{tsigklifis2017asymptotic} and \cite{Lebbal.etal.2022B}, and the study of oscillating flow in a channel between a flexible wall and a rigid wall undergoing an in-plane pulsating motion by \cite{thaokar2004stability}. Nevertheless, the main focus of these works is on the stabilizing or destabilizing influence of the modulation of the high-Reynolds-number mean flow, in contrast to the present work, in which the time-average of the basic oscillatory motion is zero.

The remainder of the paper is organized as follows. In \S\ref{sec:form}, we describe the problem configuration,
derive the basic oscillatory flow, and apply the Floquet formalism to the Navier-Stokes equations in combination
with a simple model equation for the fluid-wall interactions to derive a set of stability equations that is to be
solved numerically. The computational method employed to achieve this is set out in \S\ref{sec:nummeth}.
The results are described in detail in \S\ref{sec:results}. In particular, we focus on the marginal conditions
for the onset of instability, investigating the influence of the perturbation wavelength and the control parameters
of the problem, which are the layer widths, the Reynolds number, and the wall stiffness.
Finally, concluding remarks are given in \S\ref{sec:conc}.

%%%%%%%%%%%%%%%%%%%%%%%%%%%%%%%%%%%%%%%%%%%%%%%%%%%%%%%%%%%%%%%%%%%%%%%%%%%%%%%%%%%%%%%%%%%%%%%%%%%%%%%%%%%%%%%%%%%%%%%%
%%%%%%%%%%%%%%%%%%%%%%%%%%%%%%%%%%%%%%%%%%%%%%%%%%%%%%%%%%%%%%%%%%%%%%%%%%%%%%%%%%%%%%%%%%%%%%%%%%%%%%%%%%%%%%%%%%%%%%%%
%%%%%%%%%%%%%%%%%%%%%%%%%%%%%%%%%%%%%%%%%%%%%%%%%%%%%%%%%%%%%%%%%%%%%%%%%%%%%%%%%%%%%%%%%%%%%%%%%%%%%%%%%%%%%%%%%%%%%%%%

\section{Problem formulation}
\label{sec:form}

\subsection{Governing equations and scaling}
\label{ssec:form-goveqs}
\begin{figure}
    \centering
    \includegraphics[width=0.9\textwidth]{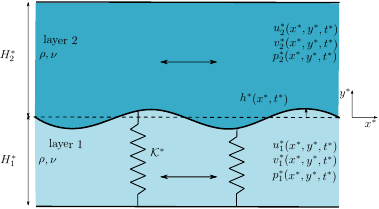}
    \caption{Schematic representation of the flow configuration.}
    \label{fig:sketch}
\end{figure}

Based on the description of cerebrospinal fluid flow in the cervical spine given in \S\ref{sec:intro}, we consider here a simplified model problem that consists of an infinitely long channel of fluid with constant density $\rho$ and kinematic viscosity $\nu$, divided longitudinally into two layers of widths $\Hs_1$ and $\Hs_2$ by an infinitesimally thin flexible wall, as depicted in figure~\ref{fig:sketch}. Both layers are subject to the same oscillatory pressure gradient in the longitudinal $\xs$-direction, $-\pfi{\ps}{\xs} = \Pils \cos(\omegas \ts)$ where $\omega^*$ denotes the angular frequency, which drives an oscillatory flow parallel to the walls. In the present analysis we are concerned with the stability of the basic oscillatory state in which the separating wall remains undeformed, and the motion in the two layers reduces to the unidirectional planar Womersley flow (see section~\ref{ssec:form-base}). In the unstable regime, the separating wall undergoes deformations that are coupled to the fluid flow. %These deformations will however remain infinitesimally small at the marginal conditions for the onset of instability, so that in the analysis the separation between layers 1 and 2 is effectively given by $\ys = 0$.

The choice of the particular fluid-structure interaction model considered here, \ie a massless undamped spring-backed plate, is motivated by the cerebrospinal fluid flow on both sides of the spinal cord, the springs mimicking the trabeculae that keep the spinal cord in place within the spinal canal (figure~\ref{fig:intro}d). The general form of the spring-backed plate-membrane equation, under the assumption of negligible tangential forces on the wall, and constraining the wall motion to the transverse $y$-direction, takes on the form \citep[see for example][]{carpenter_garrad_1985,Davies.Carpenter.1997,Shankar2002}
\begin{equation}\label{eq:solid}
    \left( m^\ast \frac{\partial^2}{\partial \ts^2} + d^\ast \frac{\partial}{\partial \ts}
+ B^\ast \frac{\partial^4}{\partial \xs^4} - T^\ast \frac{\partial^2}{\partial \xs^2} + \Ks \right) \hs (\xs,\ts)
= \left[ \ps \right]_2^1,
\end{equation}
where $\hs(\xs, \ts)$ is the transverse plate displacement, and $[\ps]_2^1$ is the pressure difference between the two sides of the membrane in the $\ys$-direction. In the model, $m^\ast$ is the membrane mass per unit of area, $d^\ast$ is the wall-damping coefficient, $B^\ast$ is the flexural ridigity, $T^\ast$ is the longitudinal tension per unit width, and $\Ks$ is the spring stiffness associated with the trabeculae. Taking into account that the spring stiffness of a single trabecular structure is on the order of $E_\text{trab} {d^\ast}^2_\text{trab}/H^\ast_\text{SSAS}$, and that there are on the order of $1/s^\ast_\text{trab}$ trabeculae per unit area, the spring stiffness coefficient $\Ks$ in equation~\eqref{eq:solid} can be estimated to be
\begin{equation}\label{Ksdef}
    \Ks \sim \frac{E_{\textup{trab}} \, d^{\ast 2}_{\textup{trab}}}{H^\ast_{\textup{SSAS}}\, s^{\ast 2}_{\textup{trab}}}.
\end{equation}
Here $E_\text{trab} \sim 12~\text{kPa}$ is the trabecular Young's modulus \citep{Jin2011}, $d^{\ast}_\text{trab} \sim 10-20\,\mu\text{m}$ the typical diameter of the trabeculae, $H^\ast_\text{SSAS} \sim 1-4\,\text{mm}$ the width of the spinal canal (in other words, the length of the trabeculae), and $s^\ast_\text{trab} \sim 500\,\mu\text{m}$ the mean distance between trabeculae (figure \ref{fig:intro}d).

%Expressing the spring stiffness in dimensionless form, it can be anticipated that, 
%
%\begin{equation}
% \K = \frac{\Ks}{\rho U^\ast \omegas} \sim 2 - 4,
%\end{equation}
%in line with those found in the stability analysis, in which the fluid-structure interaction is optimal.\\
%
%where $\Us \sim \Pils/(\rho \omegas)$ is the characteristic oscillatory velocity and $\omegas$ the oscillation frequency.

Let us now compare the order of magnitude of the other terms on the left-hand-side of equation \eqref{eq:solid} to that of the spring-stiffness term, $\Ks\hs$. In these estimations, in addition to the aforementioned properties of the trabeculae, we need the Young's modulus and Poisson ratio of the pia mater, $E_\text{pia} \sim 15\,\text{kPa}$ and $\nu_s \sim 0.5$, and its characteristic thickness, $t^\ast_\text{pia} \sim 100\,\mu\text{m}$. We also assume small deformations in the transverse direction, on the order of $h^\ast_c \sim t^\ast_\text{PIA}$, that span a longitudinal distance comparable to the stroke length of the cervical CSF motion, $l^\ast_\text{CSF} \sim 1\,\text{cm}$
\cite{carpenter_garrad_1985, howell_kozyreff_ockendon_2008}. The relative orders of magnitude of inertia, flexural rigidity and longitudinal tension with respect to the string stiffness are, respectively,
\begin{align}
    \label{eq:MK}
    \frac{O\left(m^\ast \frac{\partial^2 \hs}{\partial \ts^2}\right)}{O\left(\Ks\hs\right)}
    & \sim \frac{\rho_{\textup{pia}}t^\ast_{\textup{pia}}\omegas^2}{E_{\textup{trab}}d^{\ast 2}_{\textup{trab}}/H^\ast_{\textup{SSAS}}s^{\ast 2}_{\textup{trab}}} \sim 10^{-3}, \\
\label{eq:BK}
    \frac{O\left(B^\ast \frac{\partial^4 \hs}{\partial \xs^4}\right)}{O\left(\Ks\hs\right)}
    & \sim \frac{E_{\textup{pia}}t^{\ast 3}_{\textup{pia}}/12(1-\nu_s^2) \, l^{\ast 4}_\text{CSF}}{E_{\textup{trab}}d^{\ast 2}_{\textup{trab}}/H^\ast_{\textup{SSAS}}s^{\ast 2}_{\textup{trab}}} \sim 10^{-5}, \\
\label{eq:TK}
   \frac{O\left(T^\ast \frac{\partial^2 \hs}{\partial \xs^2}\right)}{O\left(\Ks\hs\right)}
   & \sim \frac{E_{\textup{pia}}t^\ast_{\textup{pia}} h^{\ast 2}_{\textup{c}}/(1-\nu_s^2)\, l^{\ast 4}_\text{CSF}}{E_{\textup{trab}}d^{\ast 2}_{\textup{trab}}/H^\ast_{\textup{SSAS}}\, s^{\ast 2}_{\textup{trab}}} \sim 10^{-4}.
\end{align}

Damping is also considered to be negligible compared to the spring stiffness, with $d^{\ast}\omegas/\Ks \sim 10^{-2}$ \cite{fiford2005damping}.\footnote{The reader is referred to \S\ref{sec:damping} where we briefly explore the effect of the addition of this term, which only introduces little variations in the Floquet growth rate $|\mu|$ and the critical Reynolds number $\Reyn_{cr}$ without changing the most unstable wavelength.}
It follows that in the fluid-structure interaction model~\eqref{eq:solid}, the spring stiffness is the dominating term on the left-hand-side that needs to balance the pressure difference between both sides of the wall induced by the fluid motion. Consequently, we focus our attention on the simplified model with $m^\ast = d^\ast = B^\ast = T^\ast = 0$, given by
\begin{equation}
    \Ks \hs (\xs,\ts)  = - \left[ \ps \right]_1^2.
\end{equation}
Note that this choice of fluid-wall coupling has been adopted in many biofluid mechanical problems in the literature \cite{Cani.et.al.2006, westerhof.et.al.2009, Sanchez.etal.2018}.

When considering the fluid motion, it can be anticipated that in the regime of interest, \ie near the critical conditions for the onset of instability, the thickness of the Stokes shear-wave layers near the walls is of the same order of magnitude or smaller than the channel half-widths, $\sqrt{\nu/\omegas} \lesssim \Hs_{1,2}$, while the characteristic oscillatory velocity in the bulk of the fluid layers is $\Us \sim \Pils/(\rho \omegas)$, given by the balance between the driving pressure gradient and the local acceleration. Correspondingly, the typical length scale is the stroke length $l^{\ast} \sim \Us\omegas^{-1}$. Scaling lengths with $l^{\ast}$, time with $\omegas^{-1}$, velocities with $\Us$, and pressures with $\Pils^2/(\rho \omegas^2)$, the dimensionless governing equations become (with the subindex $j = 1,2$ distinguishing between layer~1, $y < 0$, and layer~2, $y \geq 0$)
\begin{align}
    \label{eq:cont}
    \frac{\partial u_j}{\partial x} + \frac{\partial v_j}{\partial y} & = 0, \\
    \label{eq:momx}
    \frac{\partial u_j}{\partial t} + u_j \frac{\partial u_j}{\partial x}
                                    + v_j \frac{\partial u_j}{\partial y}
    & = - \frac{\partial p_j}{\partial x}
        + \frac{1}{\Reyn} \left( \frac{\partial^2 u_j}{\partial x^2}
                               + \frac{\partial^2 u_j}{\partial y^2} \right), \\
    \label{eq:momy}
    \frac{\partial v_j}{\partial t} + u_j \frac{\partial v_j}{\partial x}
                                    + v_j \frac{\partial v_j}{\partial y}
    & = - \frac{\partial p_j}{\partial y}
        + \frac{1}{\Reyn}\left( \frac{\partial^2 v_j}{\partial x^2}
                              + \frac{\partial^2 v_j}{\partial y^2} \right), \\
    \label{eq:Khp}
    \K \ms h & = p_1(y=0) - p_2(y=0).
\end{align}
The Reynolds number $\Reyn = \Us^2/(\nu\omegas) = \Us l^{\ast} / \nu = \Pils^2/(\rho^2\nu\omegas^3)$ and the dimensionless wall stiffness $\K = \Ks/\Pils = \Ks/(\rho \Us \omegas)$ are the parameters that, together with the dimensionless channel widths $H_1 = \Hs_1/l^{\ast}$ and $H_2 = \Hs_2/l^{\ast}$, govern the problem.
Note that for the flow of CSF near the spinal cord suspended by the trabecular network inside the subarachnoid space, $\K \sim 2 - 4$, as can be obtained by introducing the values of the geometric and elastic parameters of the problem in \ref{Ksdef}. In \S\ref{sec:results} we will see that the fluid-structure interaction is indeed found to be optimal for order unity values of $\K$.

The boundary conditions to be satisfied are the non-slip conditions at the channel walls and the kinematic condition at the deformable separating boundary,
\begin{equation}
    \label{eq:bc}
    \begin{cases}
    y =  H_2: & u_2 = 0, v_2 = 0,\\
    y =    h: & u_1 = u_2 = 0,  v_1 = v_2 = \pfi{h}{t},\\ 
    y = -H_1: & u_1 = 0, v_1 = 0.\\
    \end{cases}
\end{equation}
Note that when $H_1, H_2$ become very large compared to the Stokes layer thickness, of order $1/\sqrt{\Reyn}$, the velocity profiles outside the Stokes layers tend to be uniform $(\sin t, 0)$. In those cases, the Stokes layer on the nondeformable wall is so far away that it does not influence the flow near the flexible wall any longer. In the present work we focus on Reynolds numbers of order unity or larger, so that the thickness of the Stokes layer is order unity or smaller. Therefore, a pragmatic approach to study the case $H_1, H_2 \to \infty$ is to substitute the boundary conditions~\eqref{eq:bc} by
\begin{equation}
    \label{eq:bcHinf}
    \begin{cases}
    y \to  \infty: & u_2 \to \sin t, v_2 = 0,\\
    y =         h: & u_1 = u_2 = 0,  v_1 = v_2 = \pfi{h}{t},\\ 
    y \to -\infty: & u_1 \to \sin t, v_1 = 0.\\
    \end{cases}
\end{equation}

\subsection{Oscillatory base flow}
\label{ssec:form-base}

\begin{figure}
    \centering
    \includegraphics[width=\textwidth]{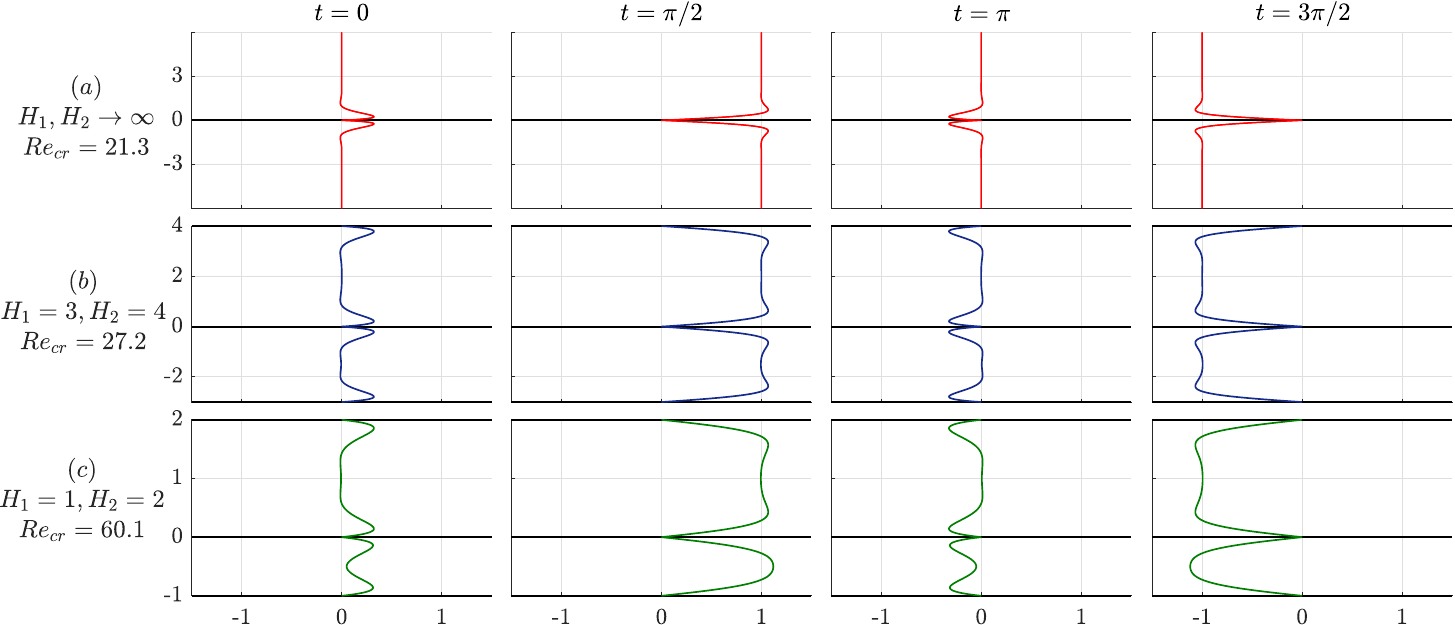}
    \caption{Oscillatory base flow at four instants of time during an oscillation cycle, for $(a)$ $H_1, H_2 \to \infty$ and $\Reyn = 21.3$, $(b)$ $H_1 = 3$, $H_2 = 4$ and $\Reyn = 27.2$, and $(c)$ $H_1 = 1$, $H_2 = 2$ and $\Reyn = 60.1$.}
    \label{fig:baseflow}
\end{figure}

The basic state about which the Floquet stability analysis is performed is the strictly parallel Womersley-like flow driven by the oscillatory pressure gradient when the separating boundary remains undeformed, given by
\begin{align}
        \ub_2(y,t) & =  \Real \left\{ \left[ \frac{1-\ue^{\beta H_2}}{\sinh(\beta H_2)}\sinh(\beta y) -\left(1-\ue^{\beta y}\right)\right]\ui\ms\ue^{\ui t} \right\},\
        \vb_2 = 0,\
        \pb_2 = -x \cos t, \\
    \ub_1(y,t) & = \Real  \left\{ \left[ \frac{1-\ue^{\beta H_1}}{\sinh(\beta H_1)}\sinh(-\beta y) -\left(1-\ue^{-\beta y}\right)\right]\ui\ms\ue^{\ui t} \right\},\
        \vb_1 = 0,\
        \pb_1 = -x \cos t, \\
    \hb & = 0, \\
\end{align}
with $\beta = (1+\ui) \sqrt{\Reyn/2}$. Notice that, since the tangential stresses are not taken into account, the base flow of both streams must be in phase for the separating surface to remain undeformed. In the limit $H_1,H_2\to\infty$, the basic state becomes
\begin{align}
    \ub_2(y,t) & = \sin t - \ue^{-y\sqrt{Re/2}} \sin(t-y\sqrt{Re/2}),\
        \bar{v}_2 = 0,\
        \bar{p}_2 = - x \cos t, \\
    \ub_1(y,t) & = \sin t - \ue^{ y\sqrt{Re/2}} \sin(t+y\sqrt{Re/2}),\
    \vb_1 = 0,\
    \pb_1 = - x \cos t, \\
    \hb   & = 0.
\end{align}
As an example, figure~\ref{fig:baseflow} shows the base flow for three different
combinations of $H_1$, $H_2$ and $\Reyn$.

\subsection{Floquet linear stability analysis}
\label{ssec:floquet}

Floquet theory \cite{iooss2012elementary} is employed to analyze the stability
of the basic oscillatory state. To that aim,
small perturbations in the form of space-periodic Floquet modes are superposed onto the basic oscillatory
state as
\begin{align}
\label{eq:floquetmodesu}
u_j & = \ub_j(y, t) + \, \uh_j(y,t) \ue^{\ui k x} \, \ue^{\sigma t}, \\
\label{eq:floquetmodesv}
v_j & = \vb_j(y, t) + \, \vh_j(y,t) \,  \ue^{\ui k x} \ue^{\sigma t}, \\
\label{eq:floquetmodesp}
p_j & = \pb_j(y, t) + \, \ph_j(y,t) \, \ue^{\ui k x} \ue^{\sigma t}, \\
\label{eq:floquetmodesh}
h   & = \hb(t) + \, \hh(t) \, \ue^{\ui k x} \ue^{\sigma t}.
\end{align}
Both, the base flow $[\ub_j(t), \ldots]$ and the amplitudes $[\uh_j(y, t), \ldots]$ are $2\pi$-periodic
functions. $k\in\mathbb{R}$ is the dimensionless perturbation wave number, and $\sigma\in\mathbb{C}$ is the
Floquet exponent. Note that the Floquet multiplier $\mu = \ue^{2\pi \sigma}$ is introduced such that
$\ue^{\sigma t} = \mu^{t/(2\pi)}$. Substitution of \eqref{eq:floquetmodesu}--\eqref{eq:floquetmodesh} into
\eqref{eq:cont}--\eqref{eq:bcHinf} yields
\begin{align}
\ui k\uh_j+\D\vh_j & = 0,\\
\left(\partial_t+\sigma\right)\uh_j+\ui k\bar{u}_j\uh_j+\bar{u}_j'\vh_j 
    & = -\ui k\ph_j + \frac{1}{\Reyn} (\DD-k^2)\uh_j,\\
(\partial_t+\sigma)\vh_j+\ui k\bar{u}_j\vh_j
    & = -\D\ph_j + \frac{1}{\Reyn} (\DD-k^2)\vh_j,\\
   \mathcal{K} \ms \hh(x,t) & = \ph_1(y=0) - \ph_2(y=0) ,
\end{align}
subject to
\begin{equation}
    \label{eq:bcfloquet}
    \begin{cases}
    y =  H_2: & \uh_2 = 0, \vh_2 = 0,\\
    y =    h: & \uh_j + \bar{u}_j'\hh = 0,  \vh_1 = \vh_2 = (\partial_t + \sigma) \hh,\\ 
    y = -H_1: & \uh_1 = 0, \vh_1 = 0,\\
    \end{cases}
\end{equation}
where $\D$ indicates partial $y$-derivatives and $\D^n = \partial^n/\partial y^n$. The kinematic boundary conditions at the perturbed interface in the normal direction, $u_j (x, y = h, t) = 0$,  has been expanded in a Taylor series about its equilibrium value at $y = h$, i.e. 
\begin{equation}
    u_j (x,y = h,t) = \left(\uh (y = h,t) + \hh(x, t)\left.\bar{u}_j'\right|_{y = h}\right)\ue^{\ui k x} = 0,
\end{equation}
where the primes indicate $y$-derivatives of the base flow, for example $\bar{u}_j' = \partial \bar{u}_j/\partial y$.
Leveraging the time-periodicity, the amplitude functions $[\uh_j(y,t), \ldots]$ are expanded as
Fourier series $[\uh(y,t) = \sum_{n=-\infty}^{\infty} \uh_{j,n}(y)\ms\ue^{\ui n t}, \ldots]$, $n \in \mathbb{Z}$. Casting the base flow in the form
$\bar{u}_j = \bar{u}_j^{\oplus}\ue^{\ui t} + \bar{u}_j^{\ominus}\ue^{-\ui t}$,
yields the following set of coupled stability equations for $n = -\infty, \ldots, 0 , \ldots,\infty$,
\begin{align}
\label{eq:contfour}
\ui k\uh_{j,n}+\mathcal{D}\vh_{j,n} & = 0, \\
\nonumber
\left[ \sigma+\ui n-\Reyn^{-1}(\DD-k^2) \right] \uh_{j,n}
+ \bar{u}_j^{\oplus}\ui k\uh_{j,n-1} + \bar{u}_j^{\ominus}\ui k\uh_{j,n+1} & \\
\label{eq:momxfour}
+ \bar{u}_j^{\oplus\prime} \vh_{j,n-1} +  \bar{u}_j^{\ominus\prime} \vh_{j,n+1}
+ \ui k\ph_{j,n} & = 0, \\
\label{eq:momyfour}
\left[ \sigma+\ui n-Re^{-1}(\DD-k^2) \right] \vh_{j,n}
+ \bar{u}_j^{\oplus}\ui k\vh_{j,n-1} + \bar{u}_j^{\ominus}\ui k\vh_{j,n+1} + \D\ph_{j,n} & = 0, \\
\label{eq:Khpfour}
\K \hh_{n} - \ph_{1, n}(y=0) + \ph_{2, n}(y=0) & = 0,
\end{align}
subject to
\begin{equation}
    \label{eq:bcfour}
    \begin{cases}
    y =  H_2: & \uh_{2,n} = 0, \vh_{2,n} = 0,\\
    y =    h: & \uh_{j,n} + \bar{u}_j^{\oplus\prime} \hh_{n-1} + \bar{u}_j^{\ominus\prime} \hh_{n+1} = 0,  \vh_{1,n} = \vh_{2,n} = (\sigma + \ui\ms n) \hh_n,\\ 
    y = -H_1: & \uh_{1,n} = 0, \vh_{1,n} = 0.\\
    \end{cases}
\end{equation}
System~\eqref{eq:contfour}--\eqref{eq:bcfour} can be written in the form of a generalized eigenvalue
problem
\begin{equation}
    \label{eq:geneig}
    \mathcal{L} \vect{q} = \sigma \mathcal{B} \vect{q},
\end{equation}
where the eigenvector
$\vect{q} = [\ldots, \vect{q}_{-n}, \ldots, \vect{q}_{-1}, \vect{q}_0, \vect{q}_1, \ldots, \vect{q}_n, \ldots]^T$
contains the Fourier modes
$\vect{q}_n = [\uh_{1,n}, \vh_{1,n}, \ph_{1,n}, \uh_{2,n}, \vh_{2,n}, \ph_{2,n}, \hh_n]^T$
that in turn contain all perturbation quantities.

The eigenvalues $\sigma$ or, equivalently, the corresponding Floquet multipliers $\mu = \ue^{2\pi\sigma}$,
determine the linear stability of the two-layer system.
When, for a certain combination of governing parameters $(\K, \Reyn, H_1, H_2)$ and for a certain wavenumber $k$,
all $|\mu| < 1$, the system is linearly stable to all perturbations of wavelength $2\pi/k$.
In the present work, we are concerned with obtaining the conditions for marginal stability, \ie the values of
$(\K, \Reyn, H_1, H_2)$ and $k$ for which a single value of $\mu$ crosses the unit circle $|\mu| = 1$.

%%%%%%%%%%%%%%%%%%%%%%%%%%%%%%%%%%%%%%%%%%%%%%%%%%%%%%%%%%%%%%%%%%%%%%%%%%%%%%%%%%%%%%%%%%%%%%%%%%%%%%%%%%%%%%%%%%%%%%%%
%%%%%%%%%%%%%%%%%%%%%%%%%%%%%%%%%%%%%%%%%%%%%%%%%%%%%%%%%%%%%%%%%%%%%%%%%%%%%%%%%%%%%%%%%%%%%%%%%%%%%%%%%%%%%%%%%%%%%%%%
%%%%%%%%%%%%%%%%%%%%%%%%%%%%%%%%%%%%%%%%%%%%%%%%%%%%%%%%%%%%%%%%%%%%%%%%%%%%%%%%%%%%%%%%%%%%%%%%%%%%%%%%%%%%%%%%%%%%%%%%

\begin{figure}
\includegraphics[width=0.99\textwidth]{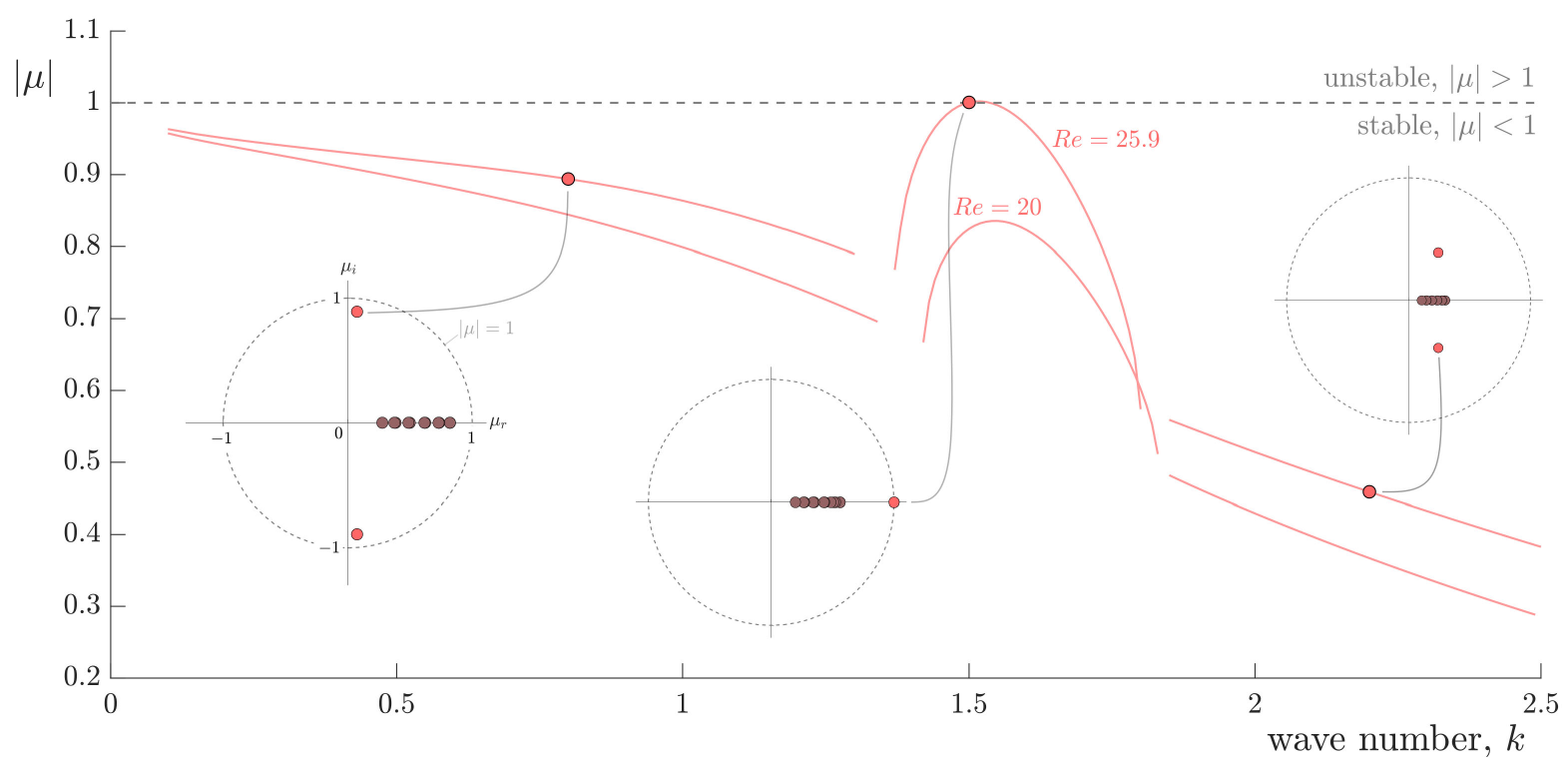}
\caption{%
Floquet growth rate $|\mu|$ as a function of the perturbation wave number $k$,
for two values of the Reynolds number, $\Reyn = 20$ and $\Reyn = 25.9$, and
$\mathcal{K} = 1$, $H_1 = H_2 \rightarrow \infty$. Insets show the spectra of Floquet
multipliers $\mu$ for three different wave numbers.
The critical conditions correspond to $\Reyn = 25.9$ and $k = 1.5$.}
\label{fig:K=1}
\end{figure}
\section{Numerical method}
\label{sec:nummeth}

The system of stability equations \eqref{eq:contfour}--\eqref{eq:bcfour} is solved numerically using a finite number
$N$ of Fourier modes.
A Chebyshev collocation method is used for the spatial discretization in the wall-normal ($y-$)direction, employing a
number $M_1$ of collocation points in layer~1, and $M_2$ points in layer~2.
Since the $n$-th Fourier mode is only coupled with modes $n-1$ and $n+1$, the general structure of the discretized eigenvalue problem is block-tridiagonal,
\begin{equation}
    \label{eq:geneigbloc}
    \setlength\arraycolsep{1.6pt}
    \begin{bmatrix}
       & & & & & & & & &\\ 
     & ...  &\vect{0}& \mathcal{L}_{n-1}^{n-2}&
     \mathcal{L}_{n-1}^{n-1}&
     \mathcal{L}_{n-1}^{n}& \vect{0}&
     ... & &\\
      & &  &  & & & & \\& &  ...&\vect{0}&
     \mathcal{L}_{n}^{n-1}&
     \mathcal{L}_{n}^{n}&
     \mathcal{L}_{n}^{n+1}& \vect{0}&
     ... &  \\
     & & & & & & & \\
       & & & ...&\vect{0}&
     \mathcal{L}_{n+1}^{n}&
     \mathcal{L}_{n+1}^{n+1}&
     \mathcal{L}_{n+1}^{n+2}& \vect{0}&... \\
      & & & & & & &
   \end{bmatrix}
   \begin{bmatrix}
   \vdots \\ \vect{q}_{n-1} \\ \vect{q}_n \\ \vect{q}_{n+1} \\ \vdots
   \end{bmatrix}
   = \sigma 
   \begin{bmatrix}
       & & &    & & & \\ 
     & ...  & \vect{0}& 
     \mathcal{B}_{n-1}^{n-1} &
     \vect{0} & 
     ... & &\\
      & &  &  & &  \\& &  ... & 
     \vect{0} &
     \mathcal{B}_{n}^{n} &
     \vect{0}  & 
     ... &  \\
     & & & & &  \\
       & & & ... &  
     \vect{0} &
     \mathcal{B}_{n+1}^{n+1} &
      \vect{0} &  
     ... \\
      & & & & & & &
   \end{bmatrix}
   \begin{bmatrix}
   \vdots \\ \vect{q}_{n-1} \\ \vect{q}_n \\ \vect{q}_{n+1} \\ \vdots
   \end{bmatrix}
\end{equation}
where each set of square submatrices
$\mathcal{L}_{n}^{n\pm1}$ and $\mathcal{B}_{n}^{n\pm1}$, of size $3(M_1+M_2)+1$,
contains the discretized stability equations \eqref{eq:contfour}--\eqref{eq:bcfour}.
Their rows are arranged such that the first $3M_1$ rows encode the continuity and
momentum equations \eqref{eq:contfour}--\eqref{eq:momyfour} for the fluid in layer~1,
the next $3M_2$ rows encode the corresponding equations for the fluid in layer~2,
and the last row contains the fluid-solid interaction equation \eqref{eq:Khpfour}.
The boundary conditions~\eqref{eq:bcfour} are implemented by replacing the corresponding
rows in $\mathcal{L}_{n}^{n\pm1}$ and $\mathcal{B}_{n}^{n\pm1}$.

The numerical implementation was performed in MATLAB$^\textrm{\textregistered}$. The number
of Fourier modes $N$ and the number of discretization points $M_1$, $M_2$ was increased until numerical
convergence was obtained (a varying number of points was needed for varying layer depths $H_1$ and $H_2$ and
varying Reynolds numbers $\Reyn$). For all cases, $N=10$ Fourier modes was found to be sufficient,
whereas $M_1$ and $M_2$ ranged from 48 to 128. %The correct implementation of the code was further verified by solving the stability of the Stokes layers near a rigid wall, for which excellent agreement was obtained with the results available in the literature \citep{Hall.1978, Blennerhassett.etal.2002}.

\begin{figure}
    \centering
    \includegraphics[width=0.7\textwidth]{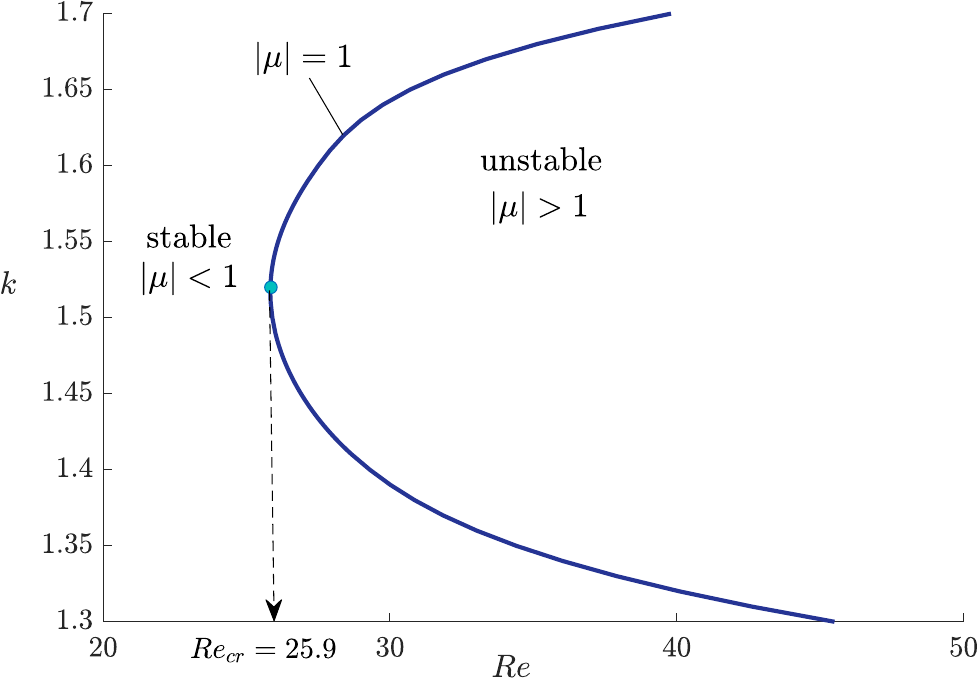}
    \caption{The curve of marginal stability corresponding to $|\mu| = 1$
    in the $(\Reyn, k)$-plane,
    for $\K = 1$ and $H_2 = H_1 = \infty$. The critical Reynolds number $\Reyn_\text{cr}$,
    indicated with a dot, is defined as the minimum $\Reyn$-value of the marginal curve.
    Note that the value $\Reyn_\text{cr} = 25.9$ corresponds to the critical condition
    shown in figure~\ref{fig:K=1}.}
    \label{fig:Re_vs_k_H1H2inf}
\end{figure}
\section{Results} \label{sec:results}
\subsection{Floquet eigenvalues and eigenmode analysis}
The numerical solution of the eigenvalue problem that determines the linear Floquet
stability of the two-layer configuration for a particular combination of canal widths $H_1$ and $H_2$,
stiffness $\K$, Reynolds number $\Reyn$, and perturbation wave number $k$, yields a spectrum of
eigenvalues~$\sigma$, or, correspondingly, a spectrum of Floquet multipliers $\mu = \ue^{2\pi\sigma}$.
The insets of Figure~\ref{fig:K=1} show such spectra for $\K = 1$, $H_1 = H_2 = \infty$,
$\Reyn = 25.9$ and three values of the wave number $k = (0.8, 1.5, 2.2)$. Whereas for
the smallest and largest
wave numbers the most unstable mode has a nonzero imaginary part (left and right inset),
there is a band of intermediate
wave numbers for which the most unstable mode is purely real (middle inset),
corresponding to perturbations of the flow field that oscillate synchronous with the base flow.
It is this branch of Floquet modes that gets most amplified when the Reynolds number increases
(compare in figure~\ref{fig:K=1}
the curve $\mu(k)$ for $\Reyn = 20$ with that for $\Reyn = 25.9$), crossing the unit circle $|\mu| = 1$ for
$k = 1.5$ and $\Reyn = 25.9$. A curve of marginal stability can be traced in the $(\Reyn-k)$-plane by
obtaining, for each value of $k$, the value of $\Reyn$ for which the most unstable mode has $|\mu| = 1$
(figure~\ref{fig:Re_vs_k_H1H2inf}). We can then define a critical Reynolds number,
$\Reyn_\text{cr}$, indicated in figure~\ref{fig:Re_vs_k_H1H2inf} with a dot,
as the minimum $\Reyn$-value of the marginal curve. Note that the
value $\Reyn_\text{cr} = 25.9$ of figure~\ref{fig:Re_vs_k_H1H2inf} corresponds of course
to the critical condition shown in figure~\ref{fig:K=1}.
The condition $(\Reyn = 25.9, \K = 1, H_1 = \infty, H_2 = \infty)$ thus lies on the marginal surface
delimiting the linearly stable and unstable regimes in the parameter space spanned by
$\Reyn, \K, H_1 \,\ \textup{and} \,\ H_2$. In section~\ref{ssec:resultsparamsweep}
different slices in this space are explored to see the individual influence of the different
physical parameters on the stability of the system.

\begin{figure}
    \includegraphics[width=1.0\textwidth]{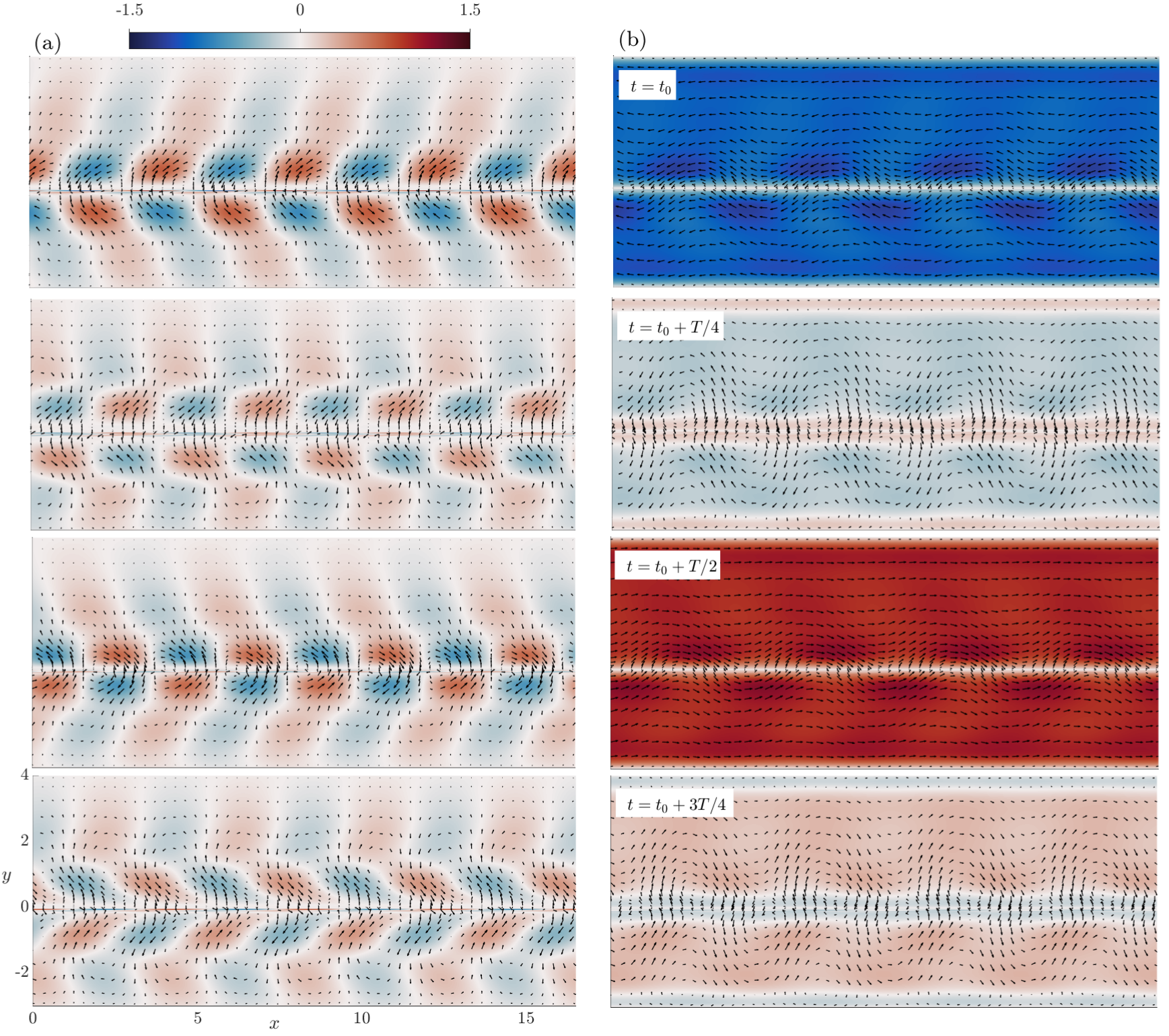}
    \caption{Time evolution of (a)  the perturbation eigenvector alone and (b), the total, perturbed, flow, computed as the
    superposition of the perturbation eigenvector and the base flow,
    $(\ub_{1,2}, \vb_{1,2}) + 0.1 (\uh_{1,2}, \vh_{1,2}) \ue^{\ui k x} \ue^{\sigma t}$,
    over the course of one oscillation cycle (with dimensionless period $T = 2\pi$),
    for the marginally unstable case
    $k = 1.52$, $\Reyn = 26$, $K = 1$, $H_1 = 3$, $H_2 = 4$.}
    \label{fig:evecs}
\end{figure}

Figure~\ref{fig:evecs} shows the time evolution of the perturbation eigenvector (a), together
with that of the total, perturbed, flow (b), computed as
the superposition of the perturbation eigenvector and the base flow,
$(\ub_{1,2}, \vb_{1,2}) + 0.1 (\uh_{1,2}, \vh_{1,2}) \ue^{\ui k x} \ue^{\sigma t}$,
over the course of one oscillation cycle (with dimensionless period $T = 2\pi$),
for the marginally unstable case $k = 1.52$, $\Reyn = 26$, $K = 1$, $H_1 = 3$, $H_2 = 4$.
Note that the perturbation eigenvector is normalized such that
$[\int_{-H_1}^{H_2} (\uh^\dag \uh + \vh^\dag \vh) \ud y]^{1/2} = 1$, where here $\dag$ indicates the complex conjugate.
Vectors indicate the direction of the flow and color represents the magnitude of the velocity perturbation in figure~\ref{fig:evecs}~(a), and of its superposition with the base flow in figure~\ref{fig:evecs}~(b). The perturbation corrugates the separating wall in a time-periodic fashion, with an associated spatially-periodic perturbed flow field that decays in the transverse direction $y$ towards the upper and lower bounding walls. The spatial wave length is dictated by the wave number $k$ as $2\pi/k = 4.1$, being $k=1.52$ in this case.

\subsection{Influence of the canal widths $H_1$, $H_2$ and the wall stiffness $\K$ on the
onset of instability}
\label{ssec:resultsparamsweep}

\begin{figure}
    \begin{subfigure}{0.49\textwidth}
        \includegraphics[width=1\linewidth]{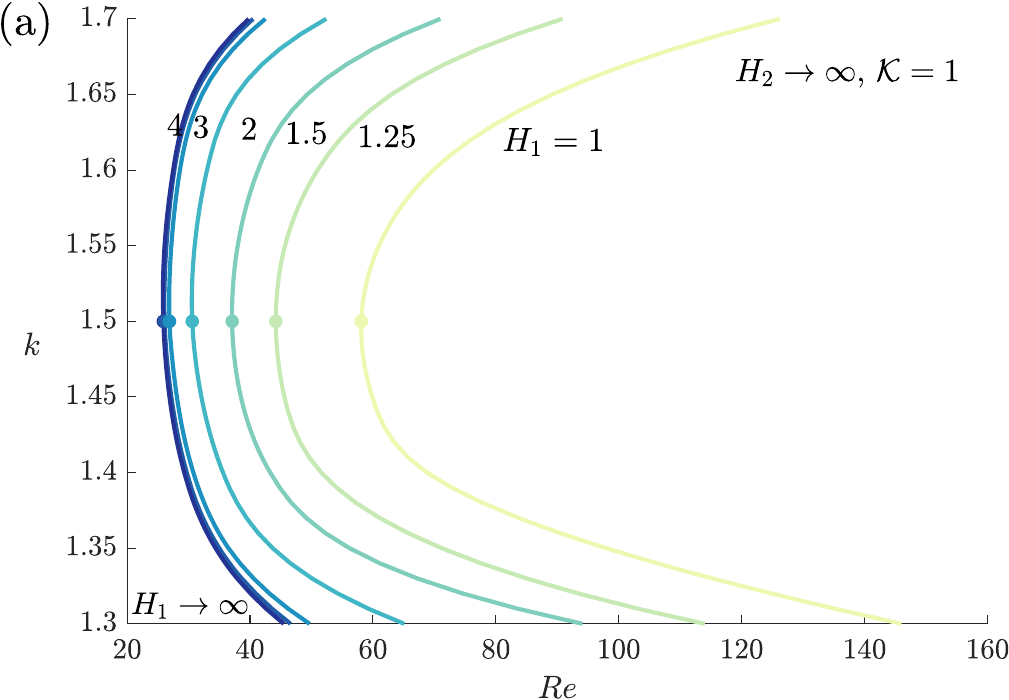}
    \end{subfigure}
    \begin{subfigure}{0.49\textwidth}
        \includegraphics[width=1\linewidth]{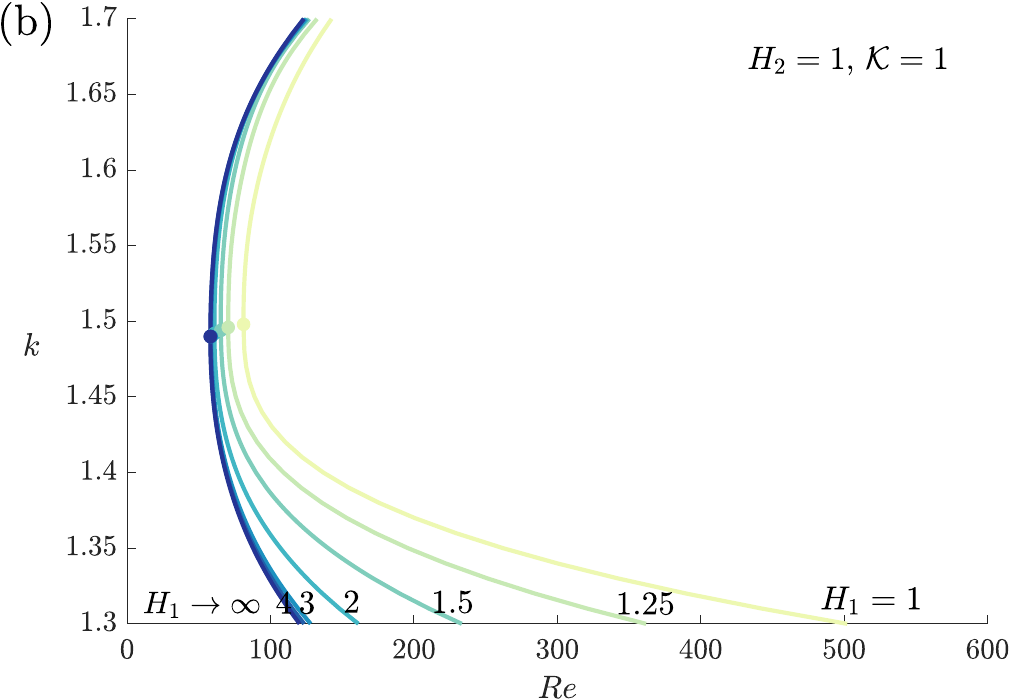}
    \end{subfigure}
     \caption{Curves of marginal stability in the $(\Reyn, k)$-plane,
     for various values of $H_1$ for the cases $(H_2 = \infty, \K = 1)$ (a)
     and $(H_2 = 1, \K = 1)$ (b). Solid dots indicate the critical Reynolds
     number $\Reyn_\text{cr}$ as the minimum $\Reyn$-value of each marginal curve.}
    \label{fig:influenceH1H2withk}
\end{figure}

\begin{figure}
    \begin{subfigure}{0.49\textwidth}
        \includegraphics[width=\linewidth]{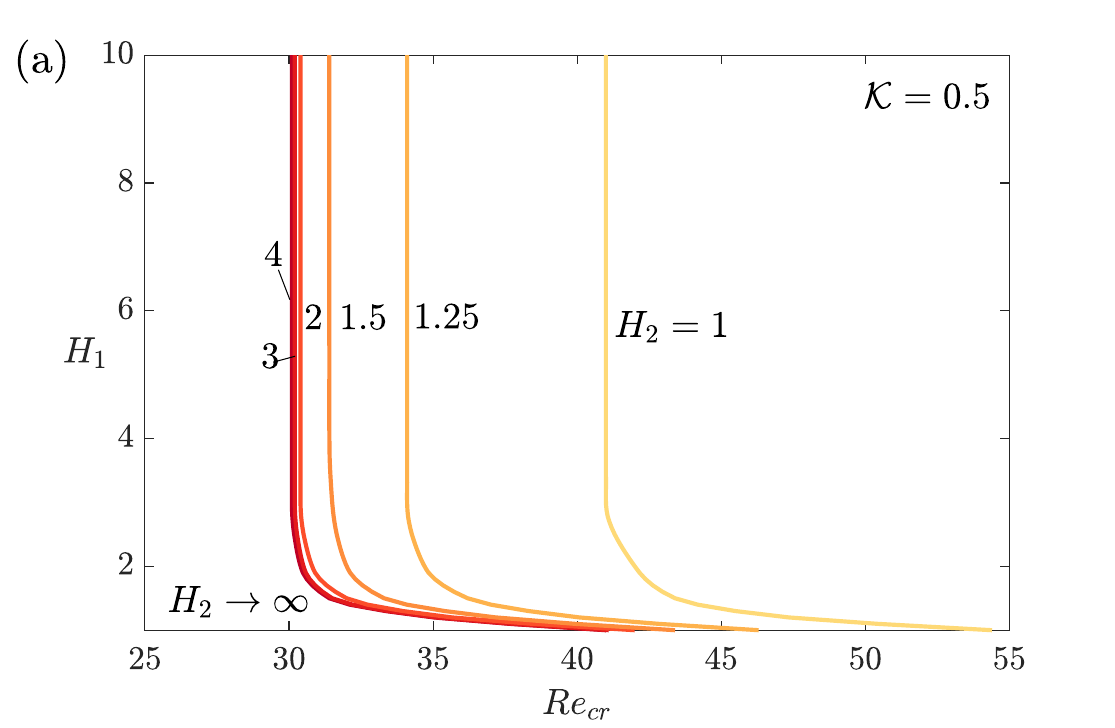}
    \end{subfigure}
    \begin{subfigure}{0.49\textwidth}
        \includegraphics[width=\linewidth]{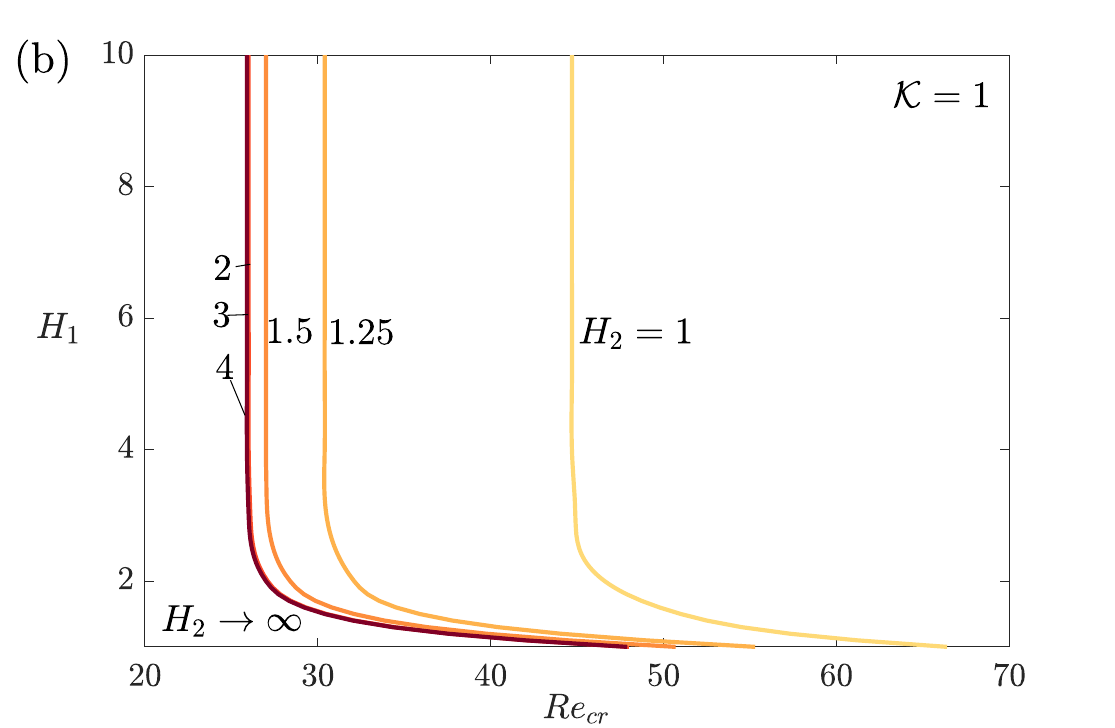}
    \end{subfigure}
    \begin{subfigure}{0.49\textwidth}
        \includegraphics[width=\linewidth]{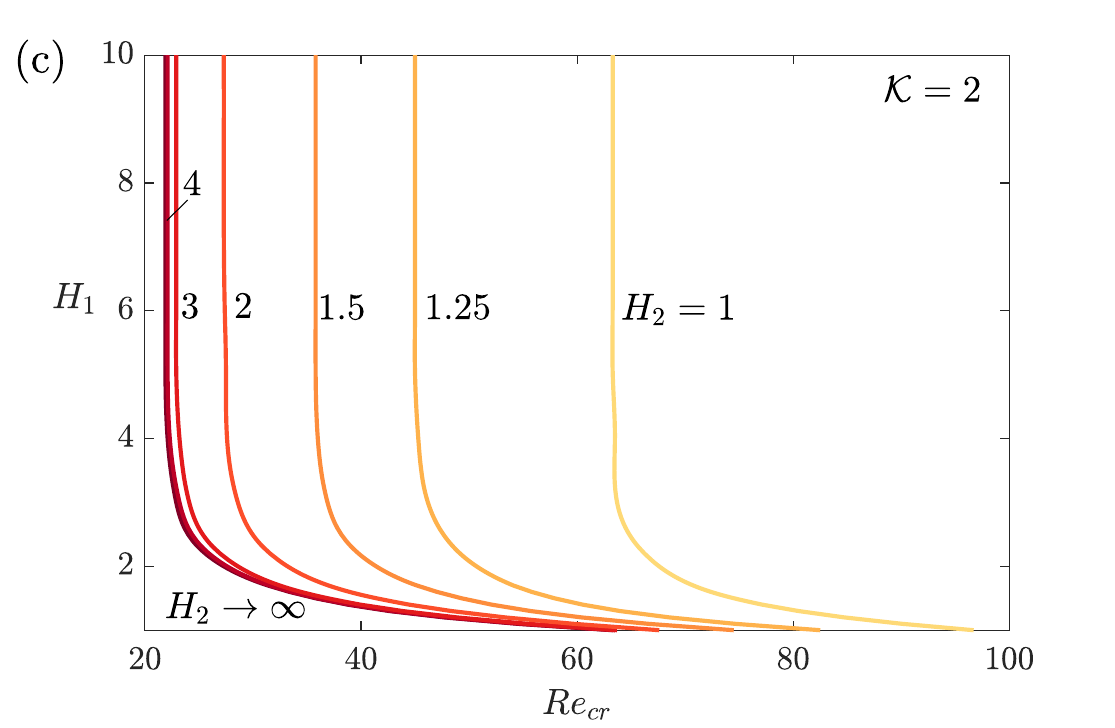}
    \end{subfigure}
     \begin{subfigure}{0.49\textwidth}
        \includegraphics[width=\linewidth]{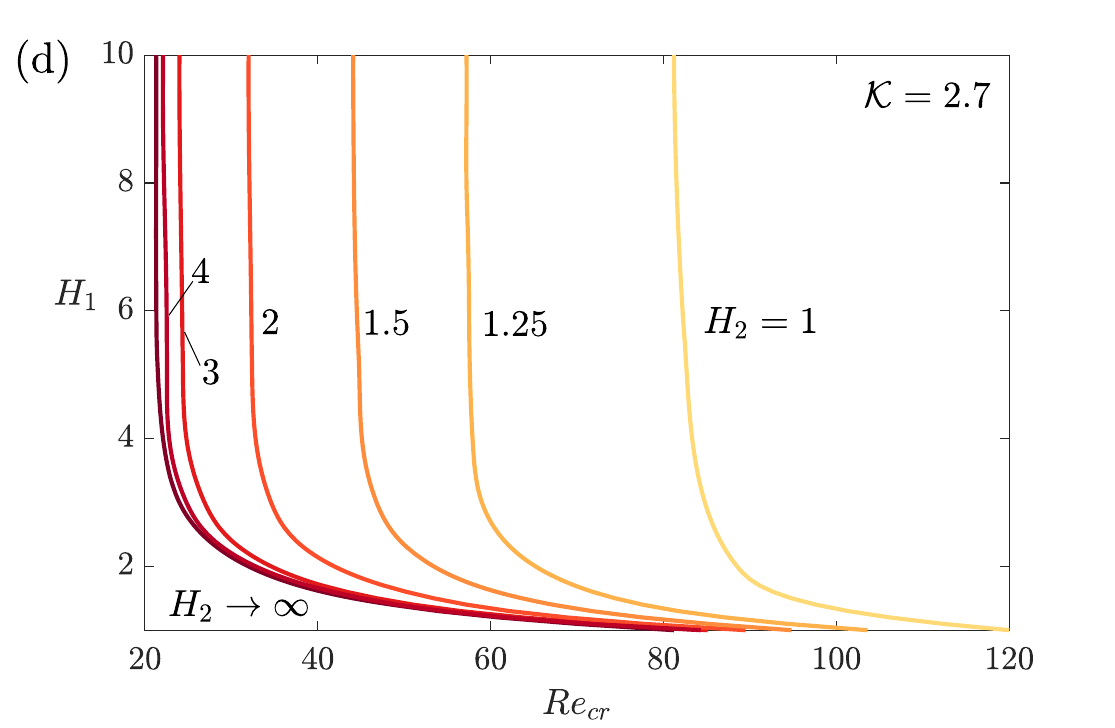}
    \end{subfigure}
    \caption{Variation of the critical Reynolds number $\Reyn_\text{cr}$ with $H_1$, for different values of $H_2$,
    and four values of $\K$: (a) $\K=0.5$, (b) $\K=1$, (c) $\K=2$ and (d) $\K=2.7$.
    Note that the wave number is approximately $k \simeq 1.94$ for all cases in (a),
    $k \simeq 1.5$ for all cases in (b), $k \simeq 0.98$ for all cases in (c), and
    $k \simeq 0.76$ for all cases in (d).}
    \label{fig:influenceH1H2condensed}
\end{figure}

A parametric analysis has been carried out to study the influence of the widths $H_1$ and $H_2$
of the two fluid layers and the wall stiffness $\K$ on the linear Floquet stability of the system.
Curves of marginal stability in the $(\Reyn, k)$-plane for different values of $H_1$ are given for
two illustrative cases: $H_2 \rightarrow \infty$, $\K = 1$  (figure~\ref{fig:influenceH1H2withk} a),
and $H_2 = 1$, $\K = 1$ (figure~\ref{fig:influenceH1H2withk} b). Reducing the canal width has a
stabilizing effect, increasing the Reynolds number associated with the onset of instability over
the entire range of wave numbers studied here. The most unstable wave number, corresponding to the
critical Reynolds numbers indicated by the dots in figure~\ref{fig:influenceH1H2withk}, is not
influenced appreciably by the canal widths, remaining close to $k = 1.5$ for the value $\K = 1$ studied in figure~\ref{fig:influenceH1H2withk}, although a slight decrease of the most unstable wavenumber can be inferred in figure~\ref{fig:influenceH1H2withk} (b) as $H_1$ increases for $H_2=1$.

\begin{figure}
    \centering
    \includegraphics[width=0.40\textwidth]{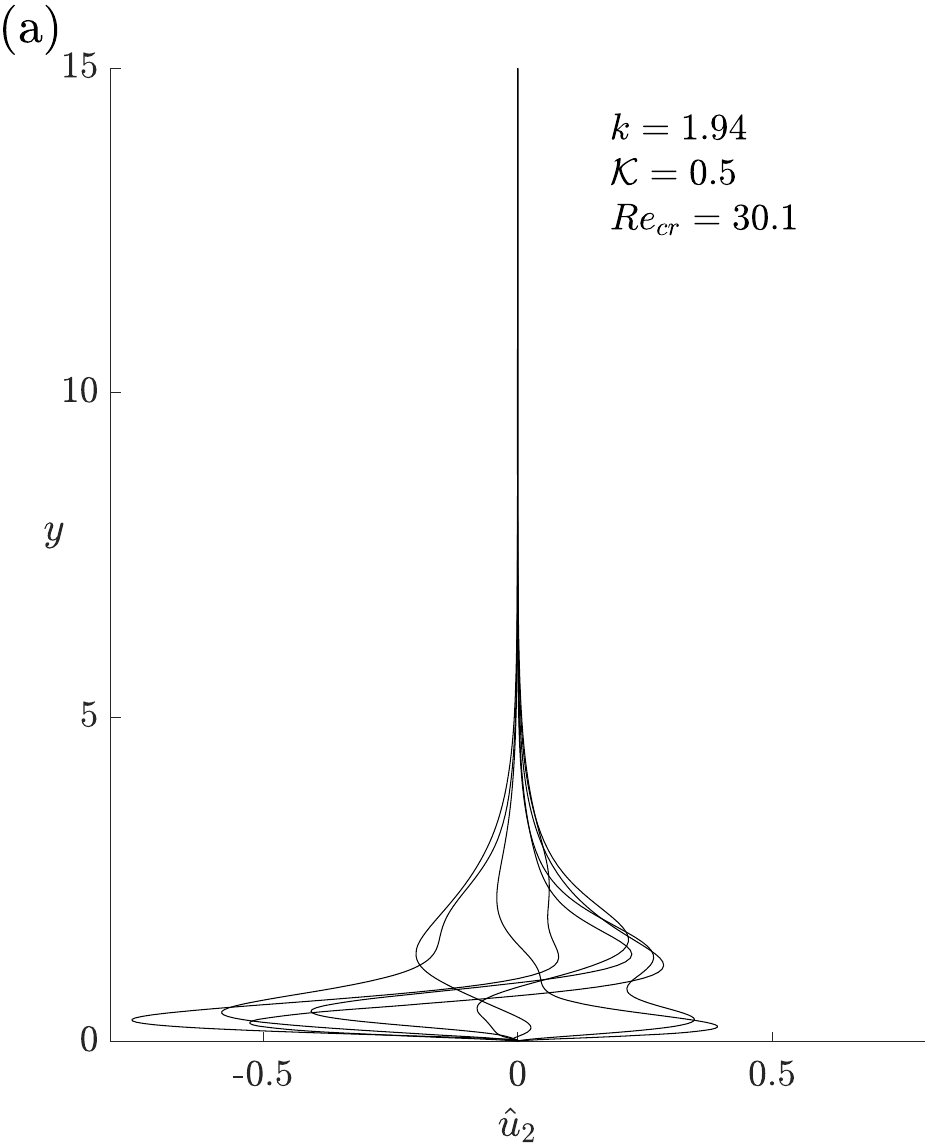}
    \includegraphics[width=0.40\textwidth]{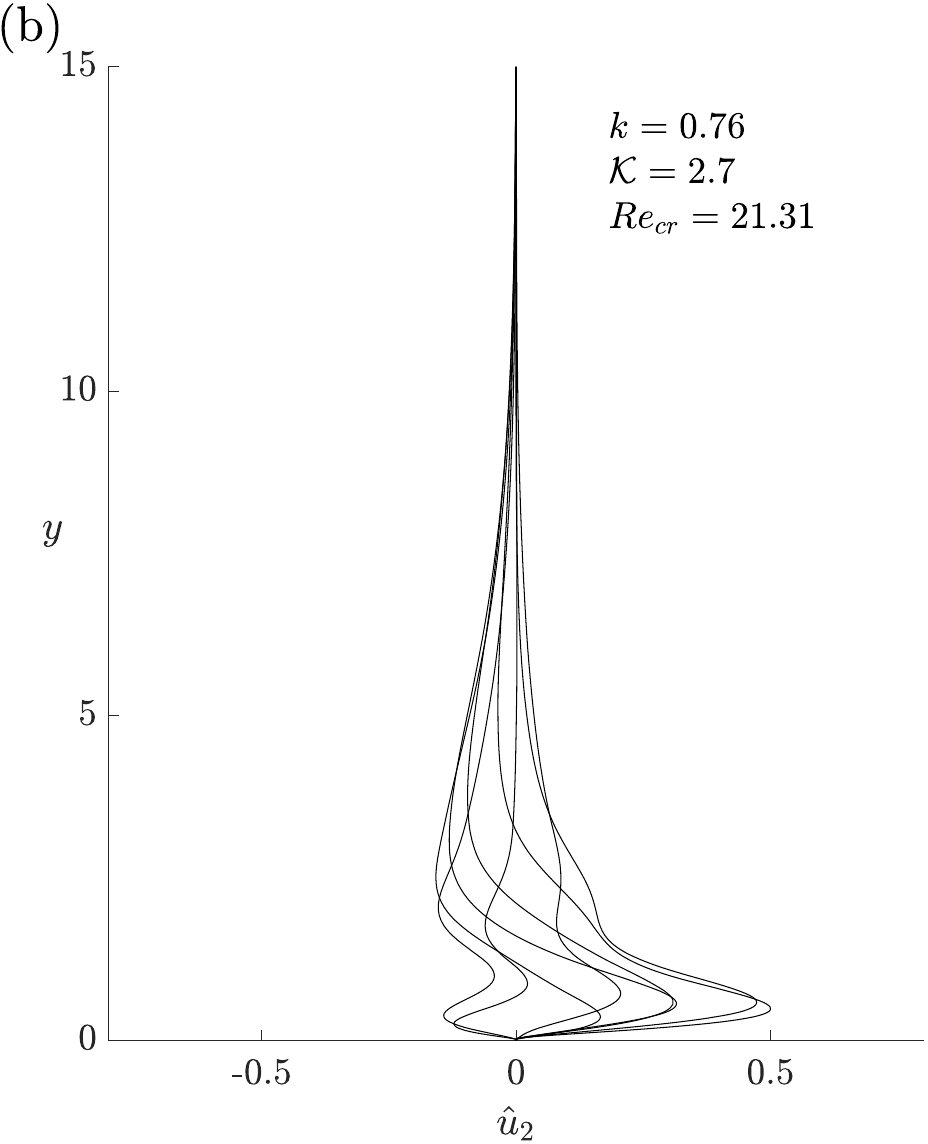}
    \caption{Velocity perturbation $\hat{u}_2(y, t)$ for two marginally stable cases for $H_1\rightarrow \infty$ and $H_2 \rightarrow \infty$:
    (a) $k = 1.94$, $\K = 0.5$ and $\Reyn = 30.1$, and
    (b) $k = 0.76$, $\K = 2.7$ and $\Reyn = 21.31$. The longer the wave length $2\pi/k$,
    the slower the transverse decay of the perturbations.}
    \label{fig:eigfundecay}
\end{figure}

Figure~\ref{fig:influenceH1H2condensed} slices the parameter space $(\Reyn, H_1, H_2, \K)$ by
showing how the critical Reynolds number varies with $H_1$, for various values of $H_2$, and
for four values of $\K$, corresponding to the four panels (a)-(d). The stabilizing influence
of decreasing the canal widths $H_1$ and $H_2$ becomes more pronounced the thinner the canals
become. This is indicated both by the slope of the marginal curves near $H_1 = 1$ and by the
distance between the different marginal curves for different $H_2$. At this point, it is worth
pointing out that, because of the symmetry of the configuration, $H_1$ and $H_2$ are
completely interchangeable in the analysis. The stabilizing
effect of confinement originates from the imposed attenuation of the velocity perturbations at the
lateral walls (boundary condition~\ref{eq:bcfour}). When unconfined ($H_1 \rightarrow \infty$
and/or $H_2 \rightarrow \infty$), the eigenfunction decays exponentially in the transverse direction
over a distance that scales with the perturbation wave length $2\pi/k$
(stemming from the $\mathcal{D}^2 - k^2$ operator in the stability
equations~\ref{eq:momxfour}--\ref{eq:momyfour}). This is illustrated
in figure~\ref{fig:eigfundecay}, which shows the velocity perturbation $\hat{u}_2(y, t)$ at 8 instants
of time over the course of one cycle for two critical cases, one case with a large wave number
($k = 1.94$, $\K = 0.5$, $\Reyn = 30.1$, figure~\ref{fig:eigfundecay} a), and another one with
a small wave number ($k = 0.76$, $\K = 2.7$, $\Reyn = 21.31$, figure~\ref{fig:eigfundecay} b). As commented previously, when changing
$H_1$ or $H_2$, the most unstable wave number, corresponding to $\Reyn_\text{cr}$, hardly changes.
Therefore, for smaller wave numbers (larger wave lengths), the transverse extent of the
eigenfunctions being larger leads to the effect that decreasing $H_1$ and/or $H_2$ on the critical Reynolds number is  noticed earlier (larger $H_1$ or $H_2$). On the contrary,
for larger wave numbers (smaller wave lengths), the eigenfunctions extend over a smaller distance,
and the confinement is noticed later (smaller $H_1$ or $H_2$).
This also explains why the influence of $H_1$ and $H_2$ occurs at different rates
for the four panels of figure~\ref{fig:influenceH1H2condensed}. Each panel corresponds to a
different value of $\K$ and has a different associated wave number $k$.

\begin{figure}
    \centering
    \includegraphics[width=0.8\linewidth]{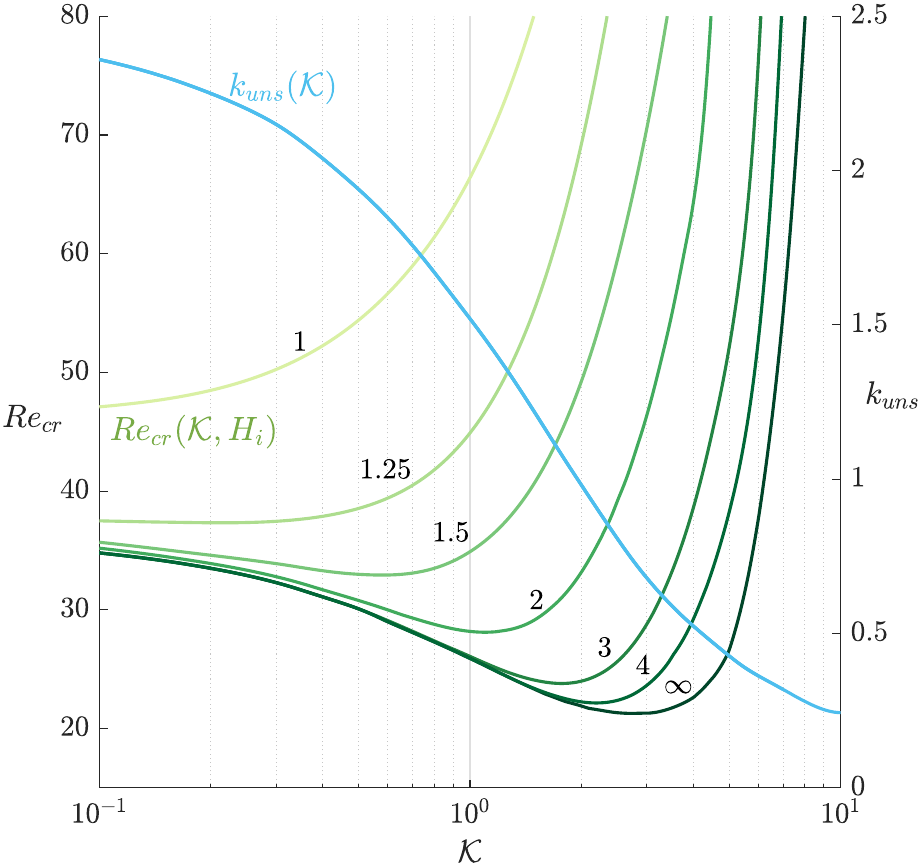}
    \caption{Critical Reynolds number $\Reyn_\text{cr}$ and corresponding wave number
    $k_\text{uns}$ as a function of the wall stiffness~$\K$ in symmetric channel configurations with $H_1 = H_2= 1, 1.25, 1.5, 2, 3, 4, \infty$. %The red dotted line indicates the value of $\K$ above which no unstable configurations could be found.
    }
    \label{fig:Figure9}
\end{figure}

The occurrence of the instability and its associated wave number strongly depend
on the value of the dimensionless stiffness $\K$.
Figure \ref{fig:Figure9} shows, in blue, how the most unstable wave number $k_\text{uns}$ decreases
with increasing stiffness $\K$ for symmetric channel configurations ($H_1 = H_2$).
Note that, in agreement with the results presented before, in the range $1 \leq H_1=H_2 < \infty$
studied here, $k_\text{uns}(\K)$ is essentially independent of $H_1$ and $H_2$,
and can therefore be presented in figure~\ref{fig:Figure9} as a single curve.
The family of green curves in this figure represents the dependence of
the critical Reynolds number, $\Reyn_\text{cr}$, with $\mathcal{K}$.
Except for the case $H_1 = H_2 = 1$, there exist a value of $\K$ for which the $\Reyn_\text{cr}$
is minimum. This minimum depends on the channel configuration, with the unbounded case being the most unstable, in the sense of exhibiting the lowest value of $\Reyn_\text{cr}$
($\K \simeq 3$). The dimensionless wall stiffness $\K$ can be interpreted as the
inverse of the square root of a reduced velocity, \ie
$\sqrt{1/\K} = \sqrt{\rho l^{\ast}/\Ks} / (\omegas^{-1})$,
which is the ratio of the characteristic time associated with the spring-backed wall
moving the adjacent fluid layer in the transverse direction, and the
longitudinal oscillation time of the basic fluid motion. Therefore, optimal fluid-structure
interaction can be expected for order-unity values of $\K$, in agreement with the non-monotonic
behavior of the marginal stability curves of figure~\ref{fig:Figure9}.

In the limit of vanishing wall flexibility, \ie when the spring stiffness $\K\to\infty$, the problem
configuration reduces to that of oscillatory flow in a channel with rigid walls, which, for
$H_1, H_2 \to \infty$ further reduces to the oscillatory Stokes-layer flow over a rigid wall.
The linear stability of the Stokes layer was first studied by \cite{Hall.1978}, and revisited by
\cite{Blennerhassett.etal.2002}, who determined that the onset of instability occurs when
the Reynolds number, based on the shear-layer thickness $\sqrt{2\nu/\omegas}$,
$\Reyn_\text{BB} = \Us \sqrt{2\nu/\omegas}/(2\nu) = 708$, for perturbations of wavelength
$\lambda_\text{BB} = \lambda^\ast / \sqrt{2\nu/\omegas} = 2\pi/0.38$.
In terms of the scales employed in the present analysis, this means a Reynolds number
$\Reyn = 2 \Reyn_\text{BB}^2 \simeq 10^6$ and a wavelength
$\lambda = \lambda^\ast/l^\ast = \lambda_\text{BB}/\Reyn_\text{BB} \simeq 2\pi/270$.
Thus, the conditions for which the wall deformation is optimally coupled to the oscillatory
fluid motion, found here for $\K$ of order unity and $\Reyn \sim 20-30$ lie very
far away from the critical conditions for the onset instability in the Stokes layer near a rigid wall.
Capturing the transition from the present problem to its rigid counterpart requires a different
treatment, using the shear-layer thickness instead of the stroke length as characteristic length scale,
and lies out of the scope of the present work.

\subsection{Influence of additional damping}
\label{sec:damping}
As detailed and discussed in \S\ref{ssec:form-goveqs}, the choice of the particular fluid-structure interaction model considered here, \ie the massless undamped spring-backed plate, is motivated by the cerebrospinal fluid flow on both sides of the spinal cord in subjects with syringomyelia, the springs mimicking micro-anatomical features that keep the spinal cord in place within the spinal canal. 

\begin{figure}
\includegraphics[width=0.90\textwidth]{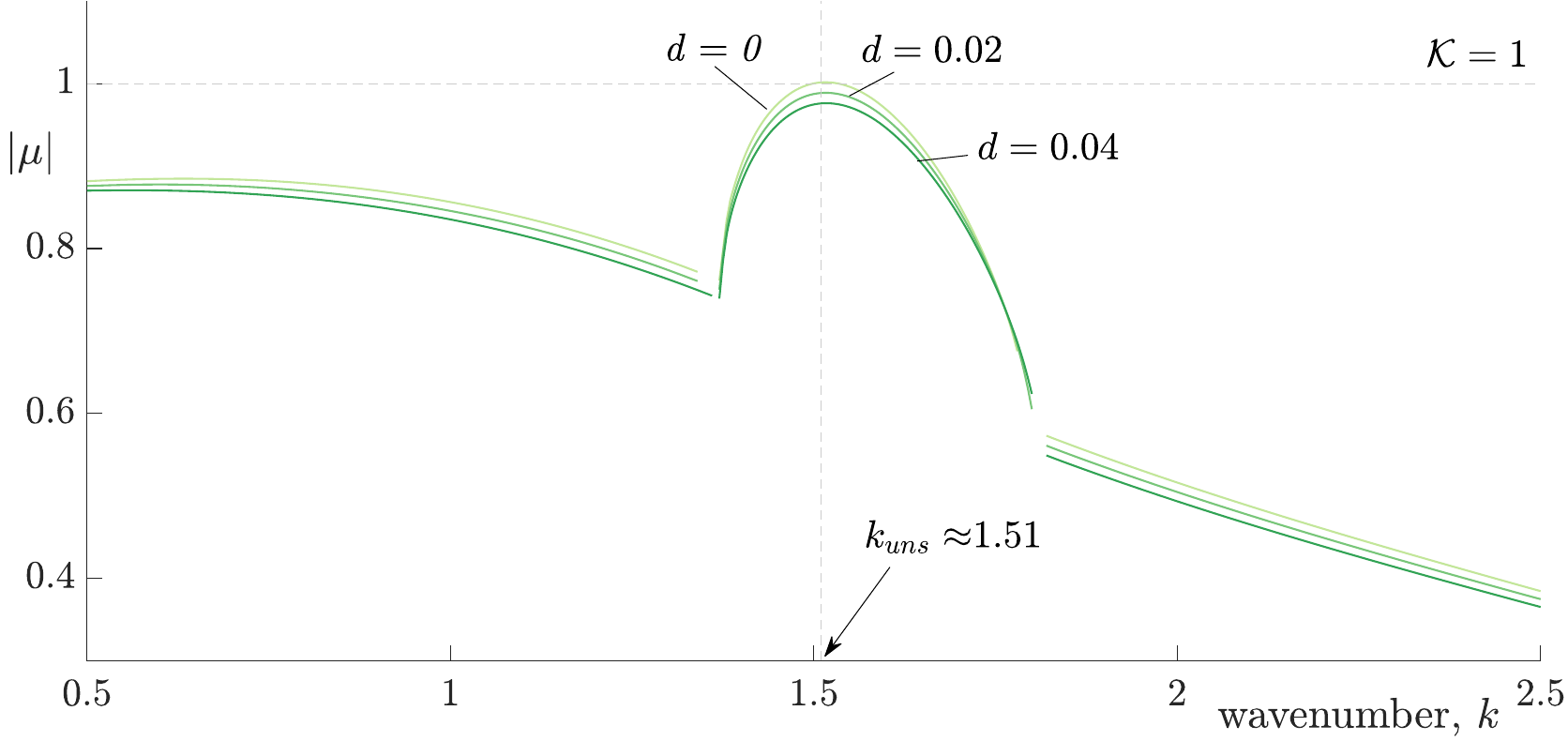}
\caption{%
Floquet growth rate $|\mu|$ as a function of the perturbation wave number $k$,
for $\Reyn = 26$, and $\mathcal{K} = 1$, $H_1 = H_2 = \infty$, when, on top of the spring stiffness,
a damping term is included in the fluid-wall interaction model,
for three values of the corresponding dimensionless damping coefficient,
$\mathscr{d} = 0$,  $\mathscr{d} = 0.02$,  $\mathscr{d} = 0.04$.}
\label{fig:Damping}
\end{figure}

Now, we briefly explore the effect of including a damping term in Eq.~\eqref{eq:Khp},
\begin{equation}
    \K \ms h + \mathscr{d}\ms{\partial h}/{\partial t} = p_1(y=0) - p_2(y=0),
\end{equation}
where $\mathscr{d} = d^\ast \omegas/\Pils$. This term is related with the amount of energy irreversibly removed from the wall. In the particular problem under consideration here, the order of the damping coefficient is estimated to be around $\mathscr{d} \sim 10^{-2}$ \cite{fiford2005damping}. Figure~\ref{fig:Damping} shows the stabilizing effect of damping on the growth rate $|\mu|$ of the perturbation (conditions similar to those of figure~\ref{fig:K=1}). Note that, the additional damping only introduces little variations in the Floquet growth rate $|\mu|$ and the critical Reynolds number $\Reyn_{cr}$, the most unstable wavelength remaining unchanged with respect to the undamped case.

%%%%%%%%%%%%%%%%%%%%%%%%%%%%%%%%%%%%%%%%%%%%%%%%%%%%%%%%%%%%%%%%%%%%%%%%%%%%%%%%%%%%%%%%%%%%%%%%%%%%%%%%%%%%%%%%%%%%%%%%
%%%%%%%%%%%%%%%%%%%%%%%%%%%%%%%%%%%%%%%%%%%%%%%%%%%%%%%%%%%%%%%%%%%%%%%%%%%%%%%%%%%%%%%%%%%%%%%%%%%%%%%%%%%%%%%%%%%%%%%%
%%%%%%%%%%%%%%%%%%%%%%%%%%%%%%%%%%%%%%%%%%%%%%%%%%%%%%%%%%%%%%%%%%%%%%%%%%%%%%%%%%%%%%%%%%%%%%%%%%%%%%%%%%%%%%%%%%%%%%%%

\section{Concluding remarks}
\label{sec:conc}

We have investigated the stability of the fluid-structure interaction problem constituted by two layers of a fluid undergoing an oscillatory motion parallel to the compliant wall that separates them. The study of this canonical configuration is inspired by the cerebrospinal fluid flow occurring in hydro-/syringomyelia, a pathology of the central nervous system characterized by the accumulation of fluid within the spinal cord, and its aim is to shed light on the underlying mechanical processes. The problem is governed by the Navier-Stokes equations for a Newtonian, incompressible, fluid, together with a constitutive equation for the compliant solid, for which a spring-backed plate equation was used, only accounting for the stiffness $\mathcal{K}$ and assuming wall-normal deformations. We investigated the linear stability of the problem by means of a Floquet analysis, focusing on the effect of the different control parameters of the problem, namely the dimensionless fluid layer widths, $H_1$ and $H_2$, the Reynolds number, $\Reyn$, and the dimensionless wall stiffness, $\K$.

Around criticality, \ie when the absolute value of the Floquet multiplier of the most unstable mode crosses the unit circle, it was found that the perturbations of the flow field oscillate synchronous with the base flow. The associated dimensionless perturbation wavenumber $k$ is of order unity, or, in other words, the spatially periodic deformations of the separating wall have a length $2\pi/k^\ast$ that is of the order of the stroke length $l^{\ast}$ of the oscillatory fluid motion. The exact value of that wavenumber depends strongly on the wall stiffness $\K$, as does the critical Reynolds number $\Reyn_\text{cr}$, underscoring the role that $\K$ plays as a reduced velocity, comparing the characteristic time of the spring-induced transverse motion with the longitudinal fluid oscillation period. Our results also reveal that reducing the channel widths has a stabilizing effect, \ie a larger Reynolds number is necessary to trigger the onset of instability over the entire range of perturbation wave numbers herein considered. In particular, such stabilizing effect has been found to become stronger as the fluid layers become thinner. The latter stems from the attenuation of the velocity perturbation close to the lateral rigid walls.

The present work aims to be a first step in understanding the complex fluid-structure interaction problem encountered in syringomyelia. It is encouraging that we find the most unstable wave lengths to be approximately $2\pi/1.5 \simeq 4$ times the stroke length $l^{\ast}$, which, with $l_\text{CSF}^{\ast} \simeq 1 \, \text{cm}$ in the spinal canal, yields $4\,\text{cm}$, comparable to the size of fluid accumulation. This may suggest that syrinx sizes lie in the range in which optimal fluid-structure interaction is achieved, and in which the motion of the intrasyringal fluid is most easily coupled with that of the cerebrospinal fluid surrounding the spinal cord. Future efforts will be focused on the validation of the results using laboratory experiments and direct numerical simulations.

\appendix
\section*{Acknowledgments}
This work was supported by the coordinated project, PID2020-115961RB-C31, PID2020-115961RB-C32 and PID2020-115961RA-C33, financed by MCIN/AEI/10.13039/501100011033, and by the Junta de Andaluc{\'i}a and European Funds, project No. P18-FR-4619.

\section*{Declaration of interest.} The authors report no conflict of interest.

\bibliographystyle{unsrtnat}

\end{document}